\documentclass[aps,twocolumn,nofootinbib,preprintnumbers,eqsecnum,superscriptaddress]{revtex4-1}

\usepackage[usenames, dvipsnames]{color}

\newcommand{\nb}[1]{\color{blue}}

\newcommand{\hl}[1]{\color{magenta}}

\usepackage[
	  pagebackref=false,
	  colorlinks=true,
      linkcolor=blue,
      urlcolor=blue,
      filecolor=black,
      citecolor=red,
      pdfstartview=FitV,
      pdftitle={},
        pdfauthor={},
        pdfsubject={},
        pdfkeywords={},
        pdfpagemode=None,
        bookmarksopen=true
      ]{hyperref}

\usepackage[normalem]{ulem}
\usepackage{amsmath}
\usepackage{enumerate}
\usepackage{amsfonts}
\usepackage{epsfig}
\usepackage{mathbbol}

\def\Tr{\mathop{\rm Tr}}
\def\tr{\mathop{\rm tr}}

\newcommand\half{{\ensuremath{\frac{1}{2}}}}
\newcommand\p{\ensuremath{\partial}}

\newcommand\vev[1]{{\ensuremath{\left\langle{#1}\right\rangle}}}

\newcommand{\be}{\begin{equation}}
\newcommand{\ee}{\end{equation}}
\newcommand{\bea}{\begin{eqnarray}}
\newcommand{\eea}{\end{eqnarray}}
\newcommand{\bega}{\begin{gather}}
\newcommand{\eega}{\end{gather}}

\newcommand{\bi}{\begin{itemize}}
\newcommand{\ei}{\end{itemize}}
\newcommand{\ben}{\begin{enumerate}}
\newcommand{\een}{\end{enumerate}}
\newcommand{\bca}{\begin{cases}}
\newcommand{\eca}{\end{cases}}
\newcommand{\bln}{\begin{align}}
\newcommand{\eln}{\end{align}}
\newcommand{\bst}{\begin{split}}
\newcommand{\est}{\end{split}}
\def\ie{\begin{equation}\begin{aligned}}
\def\fe{\end{aligned}\end{equation}}
\newcommand{\bma}{\le(\begin{matrix}}
\newcommand{\ema}{\end{matrix}\ri)}
\newcommand{\bwt}{\begin{widetext}}
\newcommand{\ewt}{\end{widetext}}

\newcommand\al{{\alpha}}
\def\b{{\beta}}
\newcommand\ep{\epsilon}
\newcommand\sig{\sigma}

\newcommand\lam{\lambda}
\newcommand\Lam{\Lambda}
\newcommand\om{\omega}

\newcommand\de{{\ensuremath{{\delta}}}}
\newcommand\De{{\ensuremath{{\Delta}}}}
\newcommand\vp{\varphi}

\newcommand\ze{\zeta}

\newcommand\da{{\dagger}}
\newcommand\nab{{\nabla}}
\newcommand\Th{{\Theta}}
\def\th{{\theta}}

\newcommand\ov{\over}
\newcommand\ha{{\half}}

\def\le{\left}
\def\ri{\right}

\newcommand\sC{{\ensuremath{{\mathcal C}}}}

\newcommand\sF{{\ensuremath{{\mathcal F}}}}

\newcommand\sL{{\ensuremath{{\mathcal L}}}}

\newcommand\sN{{\ensuremath{{\mathcal N}}}}
\newcommand\sO{{\ensuremath{{\mathcal O}}}}
\newcommand\sP{{\ensuremath{{\mathcal P}}}}

\newcommand\sR{{\mathcal R}}

\newcommand\sT{{\mathcal T}}

\newcommand\vx{{\vec x}}
\newcommand\vk{{\vec k}}

\newcommand{\rmi}{{\rm i}}
\newcommand{\rmj}{{\rm j}}

\newcommand{\heff}{{\hbar_{\rm eff}}}

\begin{document}

\title{Lectures on non-equilibrium effective field theories and fluctuating hydrodynamics}\thanks{Lectures by Hong Liu
at  ICTP SAIFR School, March 2017, Sao Paulo, and TASI, June 2017, Boulder.}
\preprint{MIT-CTP/5018}
\preprint{EFI-18-8}

\author{Paolo Glorioso}
\affiliation{Kadanoff Center for Theoretical Physics and Enrico Fermi Institute\\
University of Chicago, Chicago, IL 60637, USA}

\author{Hong Liu}
\affiliation{Center for Theoretical Physics, \\
Massachusetts
Institute of Technology,
Cambridge, MA 02139 }

\begin{abstract}

\noindent We review recent progress in developing effective field theories (EFTs) for non-equilibrium processes at finite temperature, including a new formulation of fluctuating hydrodynamics, and a new proof of the second law of thermodynamics.
There are a number of new elements in formulating EFTs for such systems. Firstly, the nature of IR variables is very different from
those of a system in equilibrium or near the vacuum.
Secondly, while all static  properties of an equilibrium system can in principle be extracted from the partition function,
 there appears no such quantity which can capture all non-equilibrium properties.
  Thirdly, non-equilibrium processes often involve {\it dissipation}, which is notoriously  difficult to deal with using an action principle. The purpose of the review is to explain how to address these issues in a pedagogic manner, with fluctuating hydrodynamics as a main example.

\end{abstract}


\maketitle

\tableofcontents

\section{Introduction}

\subsection{Effective field theory}

The goal of many-body physics is to explain and predict macroscopic phenomena.
It is, however, in general not possible to  compute macroscopic behavior  of a system directly from its microscopic description due to large number of degrees of freedom involved.  
Fortunately, for many questions of interests, 
one can separate the degrees of freedom into ``UV''  and ``IR''  variables, with characteristic spacetime scales of UV variables much smaller than those of interests. The effects of UV variables average out and one can then focus on the IR variables. Typically the relevant IR variables involve a much smaller set of degrees of freedom, and the system is greatly simplified. 


Consider, for example,
the partition function of a system at some inverse temperature $ \b$,
\be \label{par0}
Z 
= \Tr e^{-\beta H } =
 \int  D \psi \, e^{- I_0 [\psi]}  \
\ee
where $\psi$ denotes collectively all microscopic variables, 
and
$I_0$ is the microscopic (Euclidean) action.
Now imagine separating the variables into
$\{\psi\}=\{\varphi\} +\{\chi\}$
where $\vp$ and $\chi$ represent respectively UV and IR variables, and integrating out $\vp$ 
we can write the remaining integrals as
\be \label{par1}
Z = \int_\ep   D \chi \, e^{- I_{\rm EFT} [\chi; \b]}  \ .
\ee
$I_{\rm EFT}$ is the action for effective field theory (EFT)  of slow variables $\chi$. It encodes effects of UV variables and is
valid at length scales larger than some UV cutoff scale $\ep$. 

In reality integration from~\eqref{par0} to~\eqref{par1} cannot be performed directly. In fact, even the decomposition into UV and IR variables is in general not explicitly known, as the IR variables $\chi$ are often collective in nature and could
be related to microscopic variables $\psi$ in a complicated way. Nevertheless, one could often infer the physical nature of $\chi$ from experimental inputs or physical reasonings.  
We can then write down $I_{\rm EFT}$  as the most general theory of $\chi$ consistent with the symmetries (and constraints) of the system. By definition, $I_{\rm EFT}$ is nonlocal at distance scales smaller than the cutoff scale $\ep$. If we are interested in
physical processes with typical scale of variations $L \gg \ep$, $I_{\rm EFT}$ can be approximated as a local action in a derivative expansion with dimensionless expansion parameter $\ep \p_\mu \sim {\ep \ov L} \ll 1$.

The EFT approach has been tremendously successful for many problems in condensed matter and particle physics, but has been mostly formulated for systems in equilibrium or near vacuum state.
In these lectures we review some recent progress in developing EFTs for non-equilibrium processes at a finite temperature,
including a new formulation of fluctuating hydrodynamics~\cite{CGL,CGL1,Glorioso:2017lcn,Gao:2018bxz,ping}\footnote{There have been many recent activities in an action principle formulation of fluctuating hydrodynamics. See Sec.~\ref{sec:hydroeft} for other references.}  and
 a new proof of the second law of thermodynamics~\cite{GL}.
At the level of Gaussian fluctuations, these EFTs share  features with the Martin-Siggia-Rose-De Dominicis-Jansses ~\cite{msr,Dedo,janssen1} (sometimes called MSR) functional integral approaches to phenomenological  stochastic equations. However; the EFTs here  
are derived from first principles, i.e. based on symmetries and action principle, rather than from phenomenological equations. Furthermore, such EFTs can treat noise systematically at full nonlinear level.

There are a number of new elements in formulating EFTs for
non-equilibrium processes at a finite temperature. Firstly, the nature of IR variables are very different from
those for a system in equilibrium or near the vacuum.
Secondly, while all static  properties of an equilibrium system can in principle be extracted from the partition function~\eqref{par0}
 there appears no such quantity which can capture all non-equilibrium properties. Thus it is not clear a priori how to set up relevant path integrals for which an EFT can be defined. Thirdly, non-equilibrium processes often involve {\it dissipations} which are notoriously  difficult to deal with using action principle. The purpose of the review is to explain how to address these issues. For the rest of this introduction, we briefly discuss the nature  of IR variables and connection to hydrodynamics.

\subsection{Conserved quantities, local equilibrium, and hydrodynamics} \label{sec:1b}

For a system near  vacuum, the IR variables can be identified with gapless degrees of freedom: since it requires a finite amount energy to excite any gapped degrees of freedom, at low energies only gapless degrees of freedom are relevant. Now suppose we are interested in a macroscopic dynamical process of the system at a finite temperature,
will these gapless degrees of freedom remain the relevant IR variables? (Throughout this review we restrict to systems in a phase which is translationally and rotationally invariant, i.e. macroscopically a (quantum) liquid.)

The answer is no. At a finite temperature, there is now a background bath of such gapless modes. Any additional excitation will quickly be ``swallowed'' by the bath, and cannot have any direct macroscopic effect. In other words, while it takes little energy to create such an excitation, it becomes incoherent quickly. The typical time scale (and length scale)
for such an excitation to become ``incoherent'' defines the relaxation time $\tau$ (and relaxation length $\ell$).\footnote{For most systems in nature,
$\tau$ and $\ell$ are microscopic, i.e. much smaller than macroscopic spacetime scales of physical interests. In this review we will focus on such systems. Of course what one means by microscopic and macroscopic are relative. 
A somewhat extreme example is the Quark-Gluon Plasma (QGP) created at RHIC or LHC. The size of a QGP droplet is tiny, of order $10$fm, but defines the ``macroscopic scale'' of interest.  The typical relaxation length of the QGP is about $1$fm, which qualifies as being microscopic compared with the size.  For a strongly interacting system, typically $\tau \sim {1 \ov \b}$ where $\b$ is the inverse temperature.}
 In the dispersion relation of such an excitation, the frequency should have a finite imaginary part of order $1/\tau$ to reflect a lifetime of order $\tau$ and becomes ``gapped,'' thus the standard lore that finite temperature generates a gap for all excitations.

\begin{figure}[!h]
\begin{center}
\includegraphics[width=7.5cm]{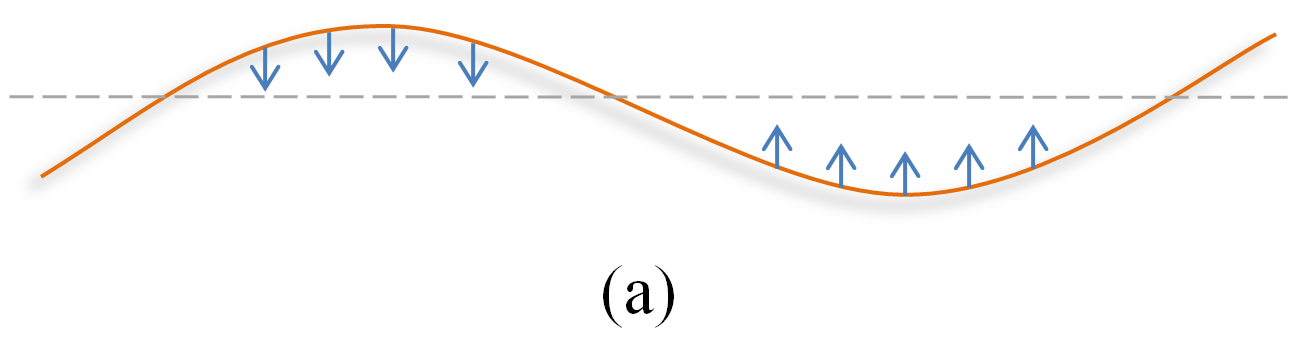}
\includegraphics[width=7.5cm]{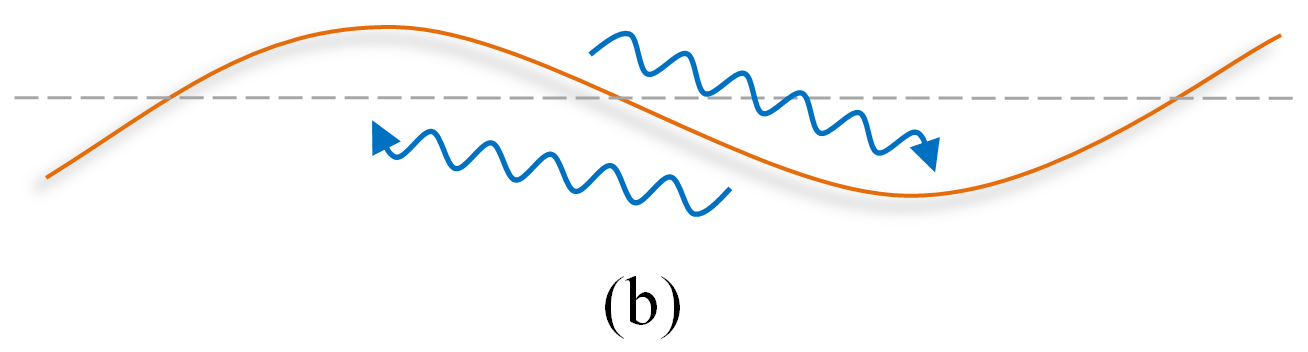}\\
\includegraphics[width=7.5cm]{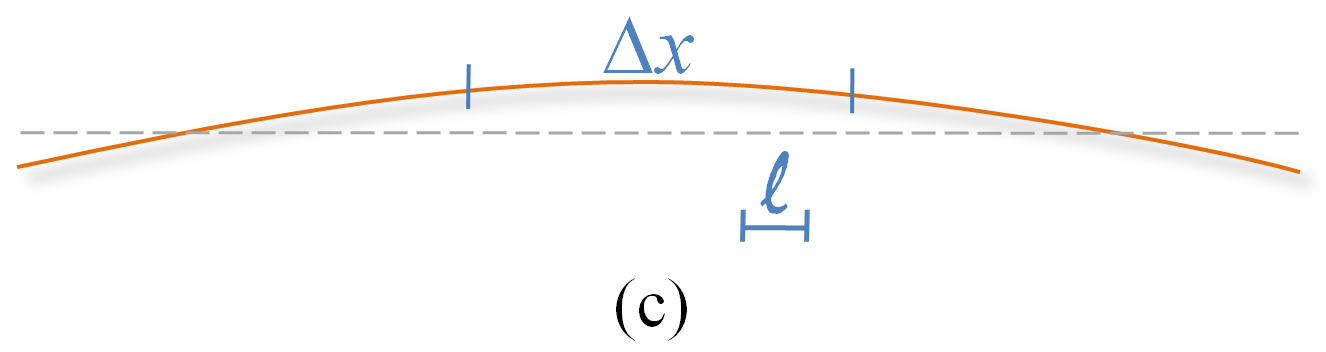}
\end{center}
\caption{Relaxation of different types of excitations. The horizontal direction is along some spatial direction. The straight dashed lines denote the global equilibrium values and the solid lines denote values of some perturbed quantities. (a) Perturbations in non-conserved quantities can relax back to equilibrium values locally--deviations separated at length scales larger than the relaxation length $\ell$ relax independently--in a time of order of the relaxation time $\tau$.
(b) Conserved quantities can only relax through transports, i.e. excesses have to be transported  to regions with deficits to achieve equilibrium. 
(c) In a spacetime region with $\ell \ll \de x \ll \lam, \tau \ll \de t \ll t_\lam$ a system can be considered as in local equilibrium
specified by the local values of conserved quantities.}
\label{fig:slow}
\end{figure}

There is, however, a caveat. Consider a long wave length perturbation of a system away from equilibrium, i.e. with wavelength $\lam \gg \ell$. Then at a time of order $\tau$, typical non-conserved quantities will have relaxed back to equilibrium.
But for a conserved quantity, which cannot be destroyed locally, relaxation back to equilibrium can only be achieved by transports.
See Fig.~\ref{fig:slow} (a) and (b).
As a result it will take time $t_{\lam} \gg \tau$ for a conserved quantity to relax.  In particular, as $\lam \to \infty,  t_\lam \to\infty$.
Thus for macroscopic physical processes involving spacetime scales much larger than $\tau$ and $\ell$, {\it the only relevant IR variables are those associated with conserved quantities,} as non-conserved quantities can be considered as in equilibrium.

More precisely, non-conserved quantities should be considered as in ``local equilibrium'' defined by the conserved quantities. To see this, consider a region of size $\de x$ satisfying $\ell \ll \de x \ll \lam$ in a time range $\tau \ll \de t \ll t_\lam$. The variations of conserved quantities in this spacetime region are small and can be considered as approximately uniform. Recall that an equilibrium state is specified by the values of conserved quantities such as energy and charge.
Non-conserved quantities in this spacetime region should then be regarded as relaxing into the local equilibrium state specified by the local values of conserved quantities.
In other words, conserved quantities are {\it low variables} which provide the background for {\it fast} relaxing
non-conserved quantities. In a non-equilibrium EFT, we integrate out fast variables and concentrate on the dynamics of slow variables.


So far we talked about generic situations. In certain special situations there can be additional non-conserved slow variables. For example, when a system is tuned to a (finite temperature) critical point, the order parameter(s) experiences critical slow-down.  Its relaxation scales become much larger than those of typical  non-conserved quantities.
Such non-conserved slow variables should also be kept in the EFT.

To summarize, for a generic system, macroscopic dynamical processes should be controlled by an EFT of slow variables associated with
conserved quantities. Given the generality of the statement, it should come as no surprise such an effective theory has in fact been widely used for a long time: it is hydrodynamics, and the variables associated with conserved quantities are usually called hydrodynamic variables.
The EFT perspective explains why hydrodynamics has been so powerful in describing so many phenomena in nature, not only in classical systems such as flow of water, patterns of weather, star and galaxy formation, but
also many exotic quantum systems 
 including the Quark-Gluon Plasma
created in heavy ion collisions at RHIC and LHC (see e.g.~\cite{Teaney:2009qa,Romatschke:2017ejr}), ultra-cold atoms (see e.g~\cite{thomas}), electron fluids in graphene~\cite{lev1,lev2,graphene1,graphene2,graphene3}, black hole physics and gauge/gravity
duality~\cite{thorne,Son:2007vk,Hubeny:2011hd}, and very recently in quantum many-body chaos~\cite{Jensen:2016pah,Saso,Lucas,Blake:2017ris}.



Despite the long and glorious history of hydrodynamics, in our opinion the potential of such a universal effective theory has
far from being fully utilized. Hydrodynamics has traditionally been formulated as a phenomenological theory in terms of
equations of motion (see Appendix~\ref{app:hydro} for a brief review).
Reformulating it from first principles as an EFT based on symmetries and action principle breaks new grounds in a number of
aspects:

\ben

\item As equations of motion, the traditional formulation of hydrodynamics cannot capture fluctuations, including both statistical and quantum fluctuations.\footnote{At linear order away from thermal equilibrium this can be partially remedied by
including some stochastic ``forces'' in the standard hydrodynamic equations, but such a manual fix does not allow systematic generalization to far-from-equilibrium situations.} Yet these fluctuations are crucial in many physical contexts, especially in far-from-equilibrium situations, including:


\ben

\item Non-equilibrium steady states and non-equilibrium phase transitions. A well known example is the onset of Rayleigh-Benard convection which is driven by hydrodynamic fluctuations,  see e.g.~\cite{Senger}.

\item Scale dependence of transport coefficients (long time tail \cite{Arnold:1997gh,Kovtun:2003vj}), which can be particularly pronounced
near phase transitions (for example, certain transport coefficients can diverge near a critical point due to hydrodynamic
fluctuations~\cite{Pomeau,hohenberg}).

\item There is an ongoing experimental program at Brookhaven National Laboratory to search for the QCD critical point using heavy ion collisions (see e.g.~\cite{Luo:2017faz}). The critical point can be probed through fluctuating properties of the QGP created, as that
close to the critical point  experiences large hydrodynamic and order parameter fluctuations (see e.g.~\cite{Plumberg:2017tvu} for a recent discussion).

\item A window into quantum gravitational fluctuations via holographic duality~\cite{CaronHuot:2009iq}.

\item In chaotic systems such as turbulent flows, tiny differences in initial states grow exponentially with time and can have macroscopic effects. Thermal fluctuations can have significant effects for turbulent flows~\cite{ruelle,eyink,mail}.

\een
A formulation of hydrodynamics based on effective action will be able to treat these problems systematically. In particular, one may be able to use
powerful field theory techniques to understand turbulence.

\item In the traditional formulation the hydrodynamic variables associated with conserved quantities are postulated based on phenomenological considerations. As such the effective theory can only apply to the regime $t_\lam \gg \tau, \lam \gg \ell$.
In~\cite{CGL}, the collective degrees of freedom associated with conserved quantities were formulated in a way {\it which
does not depend on any long wavelength expansion}. As a result the corresponding effective field theory
can in principle be valid at any scales, as far as one allows certain level of non-locality. Thus the regime of validity
of a hydrodynamic theory can be significantly extended.\footnote{There have  been very interesting recent observations of hydrodynamic attractors~\cite{Heller:2015dha,Romatschke:2017vte,Strickland:2017kux,Romatschke:2017acs}, which also suggest  that hydrodynamics can be extended beyond standard regime of validity.} In~\cite{CGL} a theory for charge diffusion which does
not use derivative expansion has been given near equilibrium, which agrees with the exact constitutive relations (again not using derivative expansion) extracted from holography~\cite{Bu:2015ame}.
A quantum hydrodynamic theory which is capable of capturing time variations of order $\tau$ has been instrumental for
a recent formulation of an effective theory for operator scrambling and quantum many-body chaos~\cite{Blake:2017ris}.

\item A formulation based on symmetries and action
principle makes transparent theoretical structures which have been
obscure in the traditional phenomenological formulation.  
For example, in the traditional formulation, one has to impose by hand the local first law and second law of thermodynamics, as well as linear Onsager relations due to underlying time reversal. It is also not clear whether these phenomenological constraints are complete.
As we will see in the EFT approach, these all follow from
 a $Z_2$ dynamical KMS symmetry, which also generalizes constraints from Onsager relations to nonlinear level~\cite{CGL}. In particular,
 the $Z_2$ symmetry together with unitarity constraints leads to a novel proof of the second law of thermodynamics for fluid systems. More recently, the effective action has also been used to clarify the
 connections among discrete symmetries, global quantum anomalies, and transports~\cite{Glorioso:2017lcn}.
New constraints which lie outside the standard entropy constraints have also been discussed in~\cite{Jensen:2018hse}.


\een




The plan of this review is as follows. In next section we discuss non-equilibrium observables of interests.
In Sec.~\ref{sec:gen} we discuss general aspects of the formulation of non-equilibrium EFTs.
In Sec.~\ref{sec:eft1}--\ref{sec:hydroeft} we discuss various examples. In particular, in Sec.~\ref{sec:eft2} we discuss
a hydrodynamic theory for diffusion and in Sec.~\ref{sec:hydroeft} fluctuating hydrodynamics for a relativistic system.
We conclude in Sec.~\ref{sec:conc} with a discussion of other generalizations.
In Appendix~\ref{app:hydro} we briefly review the standard formulation of hydrodynamics and in Appendix~\ref{app:SK} a simple example for path integral computation on a closed time path is given.

\section{Correlation functions on closed time path}\label{sec:sk}

In this section we first give a general discussion of observables in non-equilibrium systems and
then focus on properties of generating functionals for correlation functions defined on a closed time path (CTP), which are
the main observables we will focus on in this review. For standard references on closed time path or Schwinger-Keldysh formalism,  see e.g.~\cite{Chou:1984es,Niemi:1983nf,Wang:1998wg,Hubook,Kamenev,Sieberer1}.
This section will set the stage for our formulation of non-equilibrium EFTs  in later sections.

\begin{figure}[!h]
\begin{center}
\includegraphics[width=8cm]{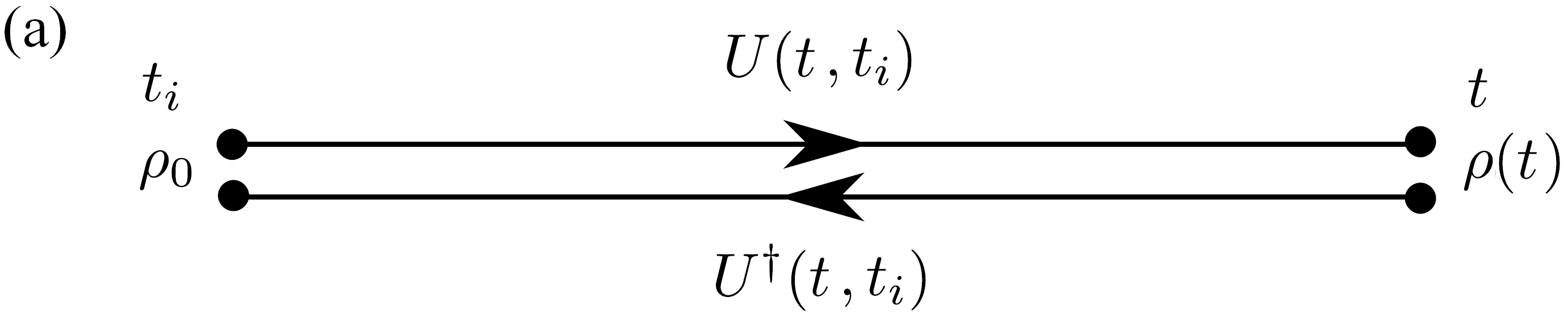} \quad
\includegraphics[width=8cm]{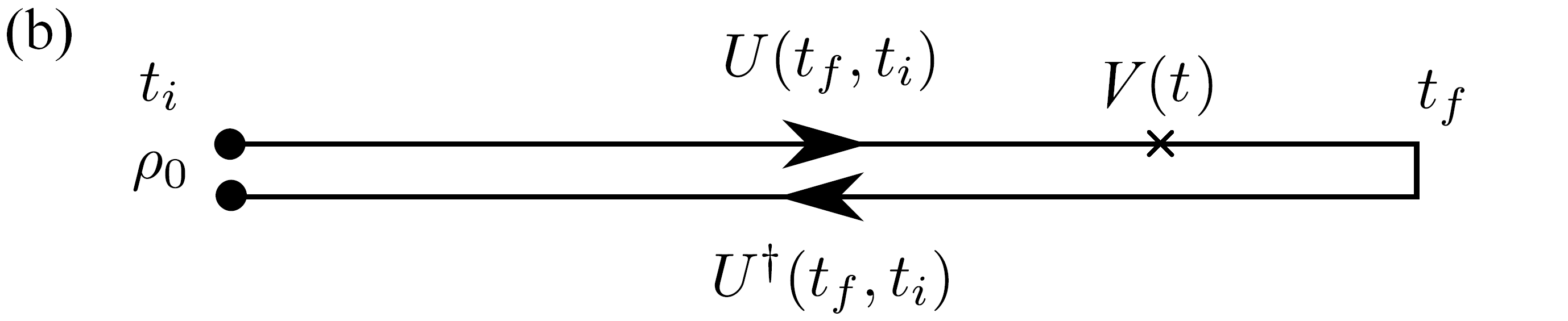}\\
\includegraphics[width=8cm]{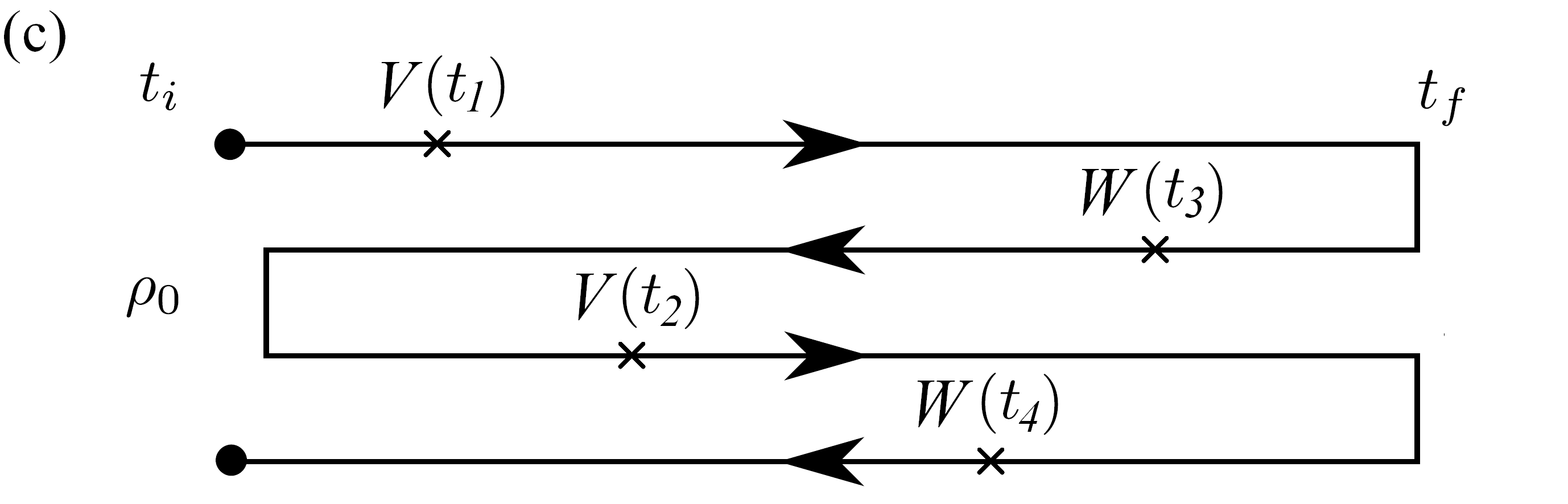} \quad
\includegraphics[width=8cm]{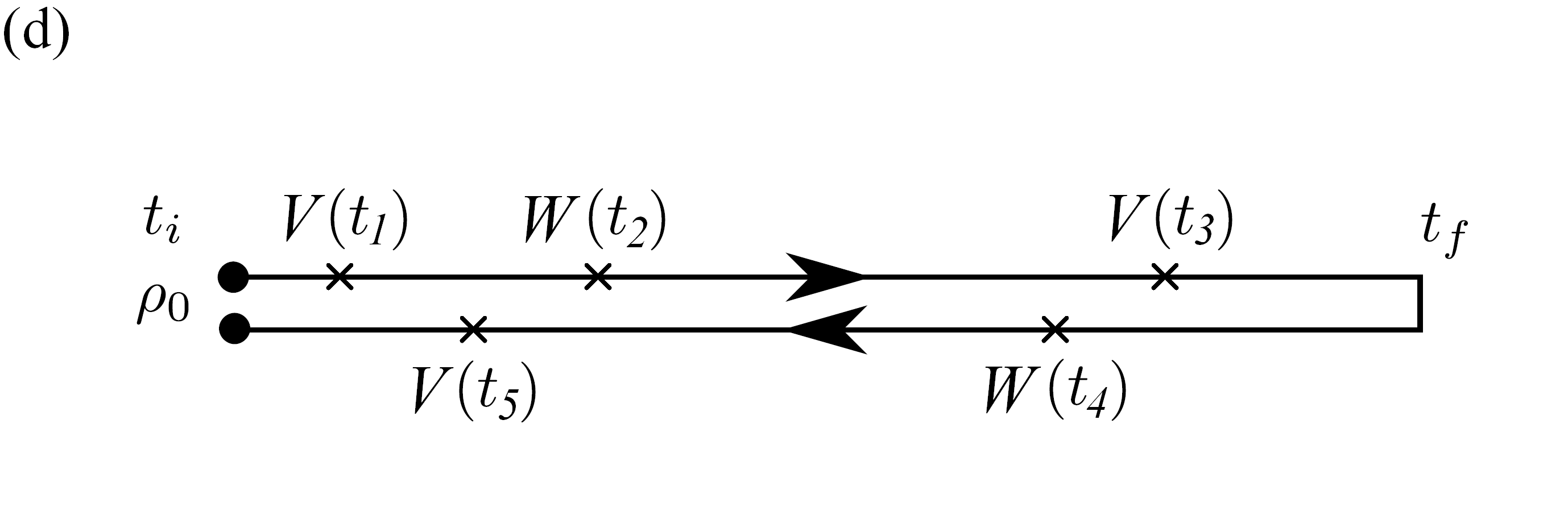}
\end{center}
\caption{(a) Path integral segments for evolution of a general initial density matrix $\rho_0$. Paths of integration are indicated by arrows.
  (b) Equation~\eqref{oo1} can be obtained by inserting $V$ at time $t$ on either segment, and joining the future ends at some time $t_f >t$.
(c) An example of the path integral contour for a general correlation function~\eqref{oo2}. 
Depending on the relative magnitudes of $t_1, t_2, t_3, \cdots$ in~\eqref{oo2}, the path integrals can have different number of segments.
Shown in figure is an example which require four segments, $\Tr(\rho_0W(t_4) V(t_2)W(t_3)V(t_1))$ with $t_1 < t_2 < t_4 < t_3$.
  To measure such an observable requires that we evolve experimental apparatus both forward and ``backward'' in time.
(d) An example of correlation function on CTP, corresponding to Eq.~(\ref{mp}).
}
 \label{fig:SK}
\end{figure}

\subsection{General non-equilibrium observables}

Consider an initial state at some time $t_i$ described by
a density matrix $\rho_0$ 
whose time evolution is given by
\be \label{enn}
\rho (t) = U(t, t_i) \rho_0 U^\da (t, t_i)\  .
\ee
Here $U(t,t_i)$ is the evolution operator from $t_i$ to $t$, and can be  expressed as a path integral from $t_i$ to $t$.
It then follows that $\rho(t)$ can be described as two path integrals,
one going forward in time from $t_i$ to $t$ and one going backward in time from $t$ to $t_i$ (see Fig.~\ref{fig:SK}a). We can probe the system with expectation values\footnote{In~\eqref{oo1}--\eqref{oo2} we have suppressed all spatial dependence, and will often do so below.}
\be \label{oo1}
\Tr (\rho (t) V) = \Tr (\rho_0 V (t)) \equiv \vev{V(t)}_{\rho_0}
\ee
which can be obtained by inserting operator $V$ at time $t$ along one of the contours
for~\eqref{enn} and then taking the trace, which in path integrals corresponds to
joining the two segments of Fig.~\ref{fig:SK}a at some $t_f > t$ as shown in Fig.~\ref{fig:SK}b.
The resulting contour is often referred to as a closed time path (CTP). One could also represent
general correlation functions
\be \label{oo2}\begin{gathered}
\vev{V (t_1) W (t_2)  X(t_3) \cdots }_{\rho_0}\\
= \Tr (\rho_0 U^\da (t_1, t_i) V U(t_1, t_2) W U (t_2, t_3) X \cdots)   \ .\end{gathered}
\ee
in terms  of path integrals as given in an example in Fig.~\ref{fig:SK}c.
Depending on the relative values of $t_1, t_2, t_3, \cdots$ in~\eqref{oo2}, the contour for the corresponding path integral
may go forward and backward in time multiple times. Due to~\eqref{enn}, the number of path integral segments is always even.
To represent a general $n$-point function one needs
 at most  $2 [{n \ov 2}]$ segments.

A simplest class of non-equilibrium observables corresponding to correlation functions obtained by inserting operators along the contour of Fig.~\ref{fig:SK}b.
This is the most general set which do not need to evolve experimental apparatus ``backward'' in time, and thus
essentially encompasses all those observables directly accessible in labs.\footnote{Observables such as those depicted in Fig.~\ref{fig:SK}c, which are called out-of-time-order correlation functions may nevertheless be measured indirectly in labs.}
  An example of correlation functions defined on a CTP is given in  Fig.~\ref{fig:SK}d, which can be written explicitly in operator form as  
\be \label{mp}\begin{gathered}
\vev{\sP V_1 (t_1) W_1 (t_2) V_1 (t_3) W_2 (t_4) V_2 (t_5)}_{\rho_0}\\
=\vev{ \tilde T (W(t_4) V (t_5)) T (V (t_1) W (t_2) V (t_3)) }_{\rho_0}\end{gathered}
\ee
where $\sP$ on the left hand side indicates that the inserted operators are path ordered with subscripts $1,2$ denoting
whether an operator is inserted on the first (i.e. upper) or the second (lower) segment. On the right hand side of~\eqref{mp} we have made explicitly that operators inserted on the first segment are time-ordered (denoted by $T$), while those on the second segment are anti-time-ordered (denoted by $\tilde T$), and the operators on the second segment always lie to the left of those on the first segment.


\subsection{Closed time path integrals and $r-a$ variables}

From now on we will restrict to correlation functions defined on a CTP contour.
In this subsection we discuss a convenient basis for them and their physical interpretations
in terms of response and fluctuation functions. 

Connected correlation functions defined on a CTP contour can be obtained from the generating functional $W$ defined
as (it is convenient to take $t_i \to -\infty, t_f \to \infty$ in Fig.~\ref{fig:SK}b)
\be\label{gen0}\begin{gathered}
e^{W [\phi_{1i}, \phi_{2i}]}
= \Tr \le[\rho_0 \sP e^{ i \int d t \, (\sO_{1i} (t) \phi_{1i} (t) - \sO_{2i} (t) \phi_{2i} (t)) }\ri]
\end{gathered}
\ee
where $\sO_i$ denote generic operators and $\phi_i$ their corresponding sources. Note that $\sO_{1i}$ and $\sO_{2i}$
are the same operator, with subscripts $1,2$ only indicating the segments of the contour in which they are inserted, while $\phi_{1i}$ and $\phi_{2i}$ are two different fields. The minus sign in the second term comes from the reversed time integration for the second (lower) segment. For definiteness we take all operators $\sO_i$ to be Hermitian and bosonic, and external sources $\phi_{1i}, \phi_{2i}$ are real.

It is useful to write~\eqref{gen0} in a few other forms

\begin{align}
&e^{W [\phi_{1i}, \phi_{2i}]} \notag \\
 & =  \Tr \le[\rho_0\! \le(\tilde T e^{- i \int d t \,  \sO_{2i} (t) \phi_{2i} (t) } \ri) \le( T e^{i  \int d t \, \sO_{1i} (t) \phi_{1i} (t) } \ri) \ri] \notag \\
&\label{pager1} \\
 &=  \Tr \le( U (+\infty, -\infty; \{\phi_{1i}\}) \rho_0 U^\da (+\infty, -\infty; \{\phi_{2i}\}) \ri)\notag
 \\
&\label{gen1} \\
&= \Tr \le[\rho_0 \sP e^{i \int d t \, (\phi_{ai} (t)  \sO_{ri} (t) +  \phi_{ri} (t) \sO_{ai} (t))}\ri] \notag \\
&\label{ra2} \ .
\end{align}

In~\eqref{gen1}, $U  (t_2, t_1; \{\phi_i\}) $ is the evolution operator of the system from $t_1$ to $t_2$ in the presence of external sources $\phi_i$.
In the last line~\eqref{ra2} we introduced the so-called $r-a$ variables
\be \begin{gathered}\label{ra1}
\phi_{ri} = \ha (\phi_{1i} + \phi_{2i}),\quad
\phi_{ai} = \phi_{1i} - \phi_{2i}, \\
\sO_{ai} = \sO_{1i} - \sO_{2i}, \quad \sO_{ri} = \ha ( \sO_{1i} + \sO_{2i})
\ .\end{gathered}
\ee

Path ordered  functions such as~\eqref{mp} are obtained by taking functional derivatives of $W$ with respect to $\phi$'s and then set the sources to zero. For example, correlation functions in the $r-a$ basis are defined as (suppressing $i,j$ indices)
\be\begin{split} \label{dfun3}
G_{\al_1 \cdots \al_n} (t_1, \cdots t_n) &\equiv {1 \ov i^{n_r}}
{\de^n W \ov \de \phi_{\bar \al_1} (t_1) \cdots
\de \phi_{\bar \al_n} (t_n)} \biggr|_{\phi_{a,r}=0}\\
 &=  i^{n_a} \vev{\sP \sO_{\al_1} (t_1) \cdots \sO_{\al_n} (t_n)} \ ,
\end{split}\ee
where $\al_1, \cdots , \al_n \in (a, r)$ and $\bar \al = r, a$ for $\al = a, r$. $n_{r,a}$ are the number of $r$ and $a$-index in $\{\al_1, \cdots, \al_n\}$ respectively ($n_a + n_r =n$). 

To get some intuition of correlation functions in the $r-a$ basis  let us  expand~\eqref{gen0} to quadratic level in external sources (i.e. consider only two-point functions) in the exponential. We then find that
\bea
&&W [\phi_{1}, \phi_{2}]  =
{i \ov 2} \int d^d x_1 d^d x_2\,\times \notag\\
&&\times\bma \phi_{1i} (x_1), \phi_{2i} (x_1)\ema
\bma G^F_{ij}  &  - i G^-_{ij}  \cr - i G^+_{ij}  & \tilde G^F_{ij}
\ema \bma \phi_{1j} (x_2) \cr \phi_{2j} (x_2) \ema  \notag\\
&&\label{ww1}\\
&& =
{i \ov 2} \int d^d x_1 d^d x_2 \,\times\notag\\
&&\times \bma \phi_{ri} (x_1), \phi_{ai} (x_1)\ema
\bma 0  &  G^A_{ij}  \cr G^R_{ij}  &  i G_{ij}^S
\ema \bma \phi_{rj} (x_2) \cr \phi_{aj} (x_2) \ema \notag\\
&& \label{ww2}
\label{z0}
 \ .
\eea
In the above equations we have used the following definitions
\be\begin{split}
G^F_{ij} (x_1, x_2)&= i \vev{T \sO_i (x_1) \sO_j (x_2)} ,\\
\tilde G^F_{ij} (x_1, x_2) &= i \vev{\tilde T \sO_i (x_1) \sO_j (x_2)} \\
G^+_{ij}(x_1,x_2) &=  \vev{\sO_i (x_1) \sO_j (x_2)}, \\
G^-_{ij}(x_1,x_2)& =  \vev{\sO_j (x_2) \sO_i (x_1)},  \\
\De_{ij} (x_1, x_2) &= \vev{[\sO_i (x_1), \sO_j (x_2) ]}, \\
\label{gsd}G_{ij}^S(x_1,x_2)& = \ha \vev{\le(\sO_i (x_1) \sO_j (x_2)+ \sO_j (x_2) \sO_i (x_1) \ri)} \\
G^R_{ij}(x_1,x_2) & = i \th (t_1 - t_2) \De_{ij} (x_1, x_2), \\
G^A_{ij}(x_1,x_2) & = - i \th (t_2 - t_1) \De_{ij} (x_1, x_2)\ ,
\end{split}\ee
where $G_R, G_A$ and $G_S$ are retarded, advanced and symmetric Green functions respectively.
See Appendix~\ref{app:SK} for an explicit evaluation of the CTP path integral at quadratic level in a simple example.

Also note that
\be\begin{split}
G^+_{ij} (x_1, x_2) &= G^-_{ji} (x_2, x_1), \\
G^R_{ij} (x_1, x_2) &=  G^A_{ji} (x_2, x_1) \ ,
\end{split}\ee
and the relations 
\be\begin{split} \label{can1}
G_F + \tilde G_F - i (G_+ + G_-) &=0 , \\
\ha \le(G_F - \tilde G_F - i (G_+ - G_-) \ri)&= G_A , \\
\ha \le(G_F - \tilde G_F + i (G_+ - G_-) \ri)&= G_R ,\\
{1 \ov 4} \le(G_F + \tilde G_F + i (G_+ - G_-) \ri)&= i G_S  \ .
\end{split}\ee

Some general remarks on~\eqref{z0} and $W$ 
 (suppress $i,j$ indices below):

\ben

\item From~\eqref{dfun3} and~\eqref{z0} we can read that
\be \begin{split}\label{yepi}
  G_{ra} (x_1, x_2) &= G_R (x_1, x_2)  ,\\
  G_{ar} (x_1, x_2) &= G_A (x_1, x_2)  ,  \\
 G_{rr} (x_1, x_2) &= G_S  (x_1, x_2) \ .
\end{split}\ee
We now see the convenience of the $r-a$ basis:  correlation functions in this basis are directly related to response ($G_R$) and fluctuation functions ($G_S$). Going beyond two-point functions, one can show that the $r-a$ correlation functions in fact correspond to the full set of nonlinear response and fluctuating functions~\cite{Chou:1984es,Wang:1998wg,bernard,peterson,Lehmann:1957zz}. More explicitly, the expectation value of an operator $\sO$
in the presence of external sources of $\phi$ can be expanded in powers of $\phi$ as
\be \begin{split}\label{dr1}
 &\vev{\sO}_\phi = \vev{\sO} +  \int d t_2 \, G_{ra} (t_1,t_2) \phi (t_2)\\
  &+ {1 \ov 2!} \int d t_2 d t_3 \, G_{raa} (t_1,t_2, t_3)  \phi (t_2) \phi (t_3) + \cdots \end{split}
\ee
with $G_{ra\cdots a}$ (with $n$ $a$'s) describing the response of $\vev{\sO}_\phi$ to external sources at $n$-th order. $G_{r \cdots r}$ (with $m$ $r$'s) is the fully symmetric $m$-point function characterizing $m$-th moment fluctuations of $\sO$, while $G_{r^m a^n}$ (with
$m$ $r$'s and $n$ $a$'s) describes the response of $m$-th moment fluctuations to external sources at $n$-th order, e.g. with $m=2$, we have
\be\begin{split}&
 \ha \vev{\{\sO (t_1), \sO(t_2)\}}_{\phi} = G_{rr} (t_1, t_2) + \\
 &\int d t_{3} \, G_{rra} (t_1, t_2 , t_{3}) \phi (t_{3}) + \cdots \ .
\label{drr1}
\end{split}\ee
Written explicitly in terms of operators $G_{r^m a^n}$ has a nested structure consisting of $n$ commutators and $m$ anti-commutators, see~\cite{Wang:1998wg,CGL} for some explicit examples.



\item In~\eqref{z0} there is no $\phi_r^2$ term, i.e.  $G_{aa} =0$, which
is due to the first identity of~\eqref{can1}. This in fact persists for general $n$-point functions. In~\eqref{gen1} taking $\phi_1 = \phi_2 = \phi$, we then find that
\be\begin{split}
 &\Tr \le( U (+\infty, -\infty; \phi) \rho_0 U^\da (+\infty, -\infty; \phi) \ri) \\
 & = \Tr (\rho_0) = 1
\end{split}\ee
and thus $W$ should  satisfy the normalization condition
\be \label{top1}
W [\phi_1 = \phi, \phi_2 = \phi] =0,\ee
or
\be
W [\phi_{a}=0 , \phi_{r}] = 0 \ .
\ee
From~\eqref{dfun3}, equation~\eqref{top1} implies that for all $n$
\be \label{3top}
G_{a\cdots a} = 0 \ .
\ee

\item Taking complex conjugate of~\eqref{gen1}
leads to a reflectivity condition
\be
\label{odd1}
W^* [\phi_1, \phi_2] = W[\phi_2,  \phi_1]  \ee
or
\be
 W^*[\phi_{a} , \phi_{r}] = W [ - \phi_{a} , \phi_{r}] \ .
\ee

\item By applying the Cauchy-Schwarz inequality to~\eqref{gen1} one finds that~\cite{GL}
\be \label{poss}
{\rm Re} \, W  [\phi_1, \phi_2] \leq 0
\ee
for arbitrary $\phi_{1,2}$. More explicitly, writing the density matrix as $\rho_0=\sum_n c_n |n\rangle\langle n|$, where $\{|n\rangle\}$ is a basis for the Hilbert space, and $0\leq c_n\leq 1$, $\sum_n c_n=1$, one finds, from (\ref{gen1}),
\be\begin{split}
\left|e^{W[\phi_{1i},\phi_{2i}]}\right|=&\Bigg|\sum_n c_n\big\langle n\big|U^\da (+\infty, -\infty; \{\phi_{2i}\})\times\\
&\times U (+\infty, -\infty; \{\phi_{1i}\})\big|n\big\rangle\Bigg|\\
&\leq  \sum_n c_n = 1\ .\end{split}\ee
Equation~\eqref{poss} can also be checked explicitly at quadratic level in~\eqref{z0} using that $G_R, G_A, G_S$ are real in coordinate space and
the definition of $G_S$ in~\eqref{gsd}.


\een

Note that~\eqref{top1}, ~\eqref{odd1}, and~\eqref{poss} all have their origin from unitarity of time evolution, i.e. from $U (+\infty, -\infty; \phi)$ being a unitary matrix.


\subsection{Thermal equilibrium and KMS conditions} \label{sec:Genth}

The discussion of the above subsection applies to any density matrix $\rho_0$. When $\rho_0$ is given by a
thermal density matrix, i.e.
\be \label{thro}
\rho_0=\frac 1{Z_0 } e^{-\beta_0H},\quad Z_0=\tr(e^{-\beta_0 H})\ ,
\ee
where $\beta_0=\frac 1{T_0}$ is the inverse temperature,  the generating functional $W$ in addition satisfies
 the so-called Kubo-Martin-Schwinger (KMS) condition~\cite{kubo57, mart59,Kadanoff}. More explicitly, using~\eqref{thro} in~\eqref{pager1}, we have
\begin{eqnarray}
&&e^{W [\phi_{1i}, \phi_{2i}]}  \notag\\
&&= {1 \ov Z_0} \Tr \bigg[e^{-(\beta_0 -\th)  H} \le(\tilde T e^{- i \int \sO_{2i} \phi_{2i}} \ri)\times\notag\\  \notag
&&\times e^{(\beta_0 - \th) H} e^{-\beta_0  H}  e^{\th  H} \le( T e^{i  \int \sO_{1i} \phi_{1i}}  \ri) e^{- \th  H}  \bigg] \\
&&=  {1 \ov Z_0} \Tr \bigg[ e^{-\beta_0  H}   \le( T e^{i  \int \sO_{1i} \phi_{1i} (t + i \th)}  \ri)\times\notag\\
&& \times \le(\tilde T e^{- i \int \sO_{2i} \phi_{2i} (t - i (\beta_0-\th))} \ri) \bigg]   \label{newfdt}\\
 &&\equiv  e^{W_T  [\phi_{1i} (t + i \th ), \phi_{2i} (t - i (\beta_0-\th)) ] } \
\label{defwt}
\end{eqnarray}
where $\th \in [0, \beta_0]$ is a constant, and in the second line we have used that  for arbitrary $a \in [-\b_0, \beta_0]$
\be \label{imts}
e^{- a H} \le(\tilde T e^{ i \int \sO (t) \phi (t)} \ri) e^{a H}
= \tilde T e^{i  \int \sO(t) \phi (t - i a)}, \quad
\ee
and similarly for the $T$ ordering factor. Equation~\eqref{imts} should be understood as being applicable under thermal
averages and its validity is a consequence of analytic properties of thermal correlation functions.
Note that the second line~\eqref{newfdt} has time ordering before anti-time ordering which is different from~\eqref{pager1}. So in the third line~\eqref{defwt}
we have introduced a new notation to denote it.

\begin{figure}[!h]
\begin{center}
\includegraphics[width=7.5cm]{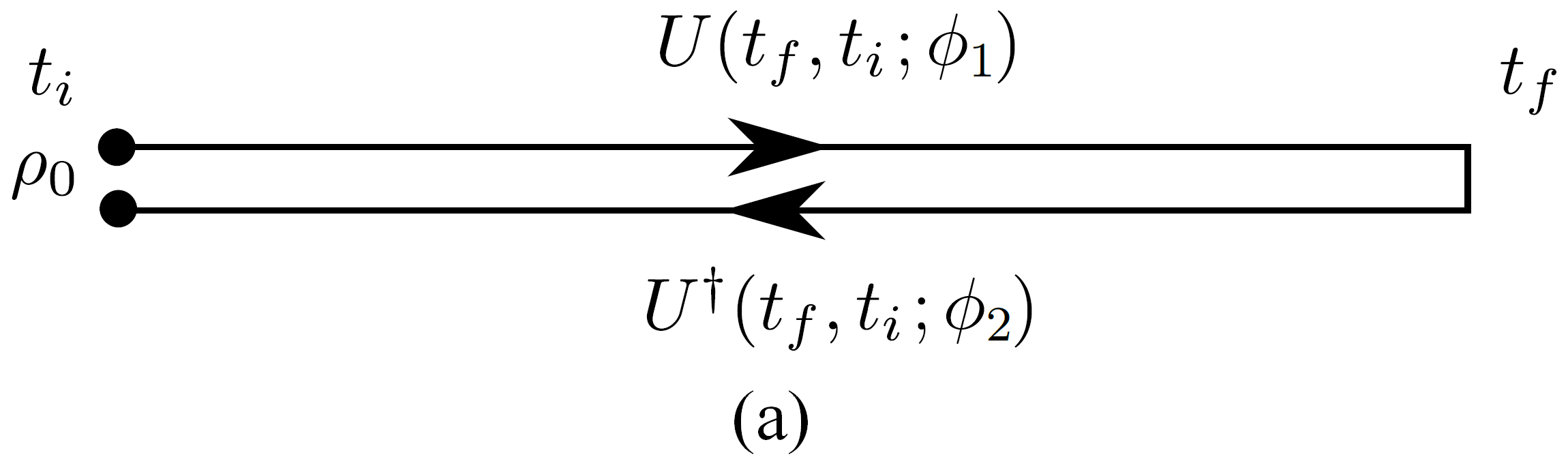}
\includegraphics[width=7.5cm]{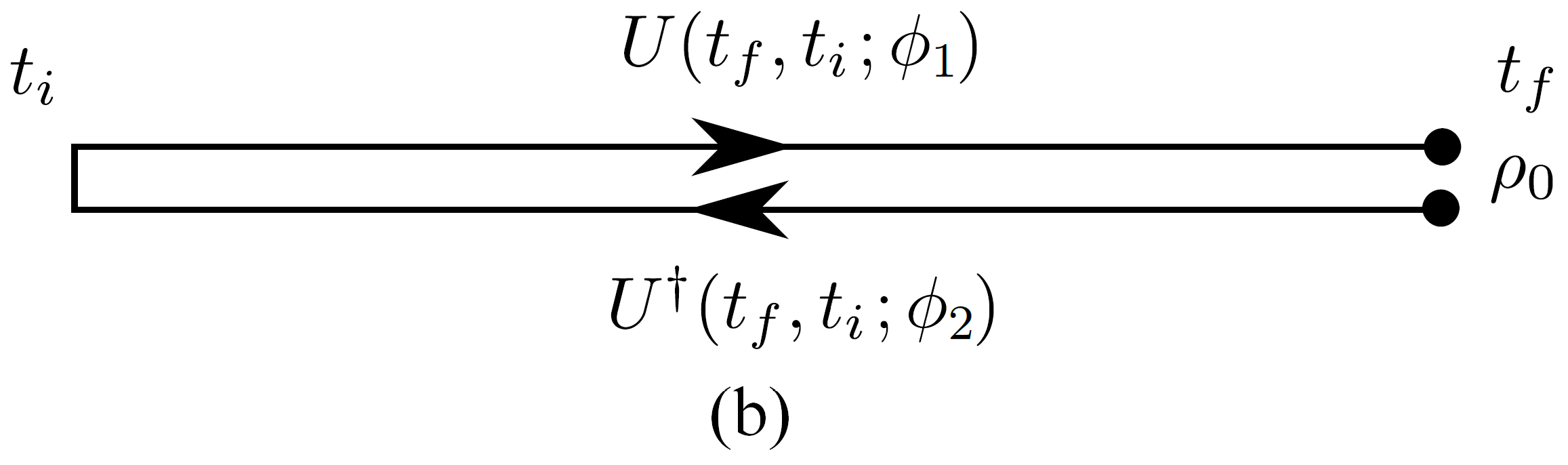}
\end{center}
\caption{(a) Integration contour corresponding to $W$. (b) Integration contour corresponding to $W_T$ as defined in~\eqref{newfdt}.}
 \label{fig:SKR}
\end{figure}

Expanding~\eqref{newfdt} to quadratic order in external sources as in~\eqref{ww1}--\eqref{ww2}, one finds that $W_T$ can also be expressed in terms
of $G_R, G_A, G_S$.  From~\eqref{defwt},
one finds  the standard fluctuation-dissipation theorem (FDT) for two-point functions
which in momentum space has the form
\be \label{fdt4}
G_{ij}^S (k) 
= \ha \coth {\beta_0 \om \ov 2} \De_{ij} (k)  \ . 
\ee
But for three-point functions and higher, the KMS condition~\eqref{newfdt}
does not by itself impose any constraints on $W$, as $W_T$ is expressed in terms of a different set of correlation functions
from $W$.  Translating~\eqref{newfdt}
to path integrals we find that $W_T$ corresponds to the generating functional for the contour indicated in Fig.~\ref{fig:SKR}, with $\rho_0$ being the final state. Hence, the KMS condition~\eqref{newfdt} relates correlation functions with $\rho_0$ as the initial state to those with $\rho_0$ as the final state. It can be readily checked that~\eqref{newfdt} is a $Z_2$ operation; when acting  twice one simply gets back $W$ itself up to an overall time translation, which does not lead to any constraint.


Now let us assume that at microscopic level the system has
an underlying discrete symmetry $\Th$ which includes time reversal, i.e. $[\Th, H] =0$.  Here  $\Th$ can be the time reversal $\sT$ itself, or any combinations of $\sC, \sP$ with $\sT$, such as $\sC\sP\sT$. Then combing $\Th$ and~\eqref{newfdt}
we can obtain a constraint on $W$~\cite{CGL}
\be
\label{1newfdt1}
W [\phi_1 (x), \phi_2 (x)] = W [\tilde \phi_1 (x), \tilde \phi_2  (x)]
\ee
where we have restored spatial dependence with $x$ denoting $x^\mu = (x^0, x^i)  = (t, \vx)$, and
\be \begin{split}\label{tiV0}
\tilde \phi_1 (x) &= \Th \phi_1 (t - i \th, \vx) , \\
\tilde  \phi_2 (x) &=  \Th \phi_2 (t + i (\beta_0 - \th), \vx )
\end{split}\ee
for arbitrary $\th \in [0, \beta_0]$.
In the above equation, the action of  $\Th$ on a spacetime tensor field $ G (x)$ should be understood as
\be \label{tg1}
\Th G (x) \equiv \eta_G G (\eta x) , \qquad \Th^2 G (x) = G (x)
\ee
where we have suppressed spacetime indices of $G$ and $\eta_G$ should be understood as a collection of phases ($\pm 1$)--one for each spacetime component for $G$. See Appendix~\ref{app:c} for how various variables transform under different choices of $\Th$. For examples, suppose $\Th = \sC\sP\sT$ and $\phi_{1,2}$ are neutral scalars with $\eta_\phi =1$,
then~\eqref{tiV0} can be written more explicitly as\footnote{Note $\Th \phi (t + z, \vx) = \phi (-t + z^*, -\vx)$ for a complex shift $z$.}
\be
\begin{split}
\tilde \phi_1 (x)& = \phi_1 (-t + i \th, - \vx ) , \\
\tilde  \phi_2 (x) &=  \phi_2 (- t - i (\beta_0 - \th), -\vx )  \ .
\label{tiV}
\end{split}
\ee
Below we will simply refer to~\eqref{1newfdt1} as the KMS condition, but it should be kept in mind it also encodes consequences of microscopic time-reversal symmetry.

We emphasize that~\eqref{1newfdt1} is fully non-perturbative in external sources. Given that the presence of finite external sources takes the system far away from the thermal equilibrium,~\eqref{1newfdt1} thus constrains the system in far-from-equilibrium
situations.

For two-point functions, equation~\eqref{1newfdt1} with~\eqref{tiV} implies, in addition to~\eqref{fdt4}
\be \label{1onsa}
G_{ij}^S (k)  = G_{ij}^S  (-k)  , \qquad G_{ij}^R (k) = G_{ji}^R (k), 
\ee
the second of which are Onsager relations. Note that the first equation implies that
\be
G_{ij}^S (k)  = 
G_{ji}^S (k)
\ee
as $G_{ij}^S$ is real in coordinate space and is Hermitian in momentum space.

\subsection{Nonlinear Onsager relations and connection to partition function}

For higher-point functions the implications of~\eqref{1newfdt1} become increasingly complicated~(see \cite{Wang:1998wg,CGL} for some examples).
In general it relates nonlinear response functions to various response-fluctuation functions and thus can be considered the
nonlinear generalizations of the FDT~\eqref{fdt4}. It was found in~\cite{CGL} that~\eqref{1newfdt1} also imposes
a set of constraints on nonlinear response functions, i.e. functions $G_{ra\cdots a}$ with only one $r$-index.
Those constraints can be separated into two classes: one class can be interpreted as corresponding to nonlinear
generalization of the Onsager relations while the other class tells us how to extract the equilibrium partition function~\eqref{par0} (in the presence of sources) from the generating functional~\eqref{gen0}.

\ben
\item Consider a system in the presence of external sources $\phi_i$ for operators $\sO_i$, with one-point functions of $\sO_i$ in the presence of sources given by $\vev{\sO_i}_\phi$. The two-point response functions in the presence of sources are  given by 
\be
G^{R}_{ij} (x, y; \phi_k (x)]= {\de \vev{\sO_{i} (x)}_\phi \ov \de \phi_{j} (y)} \
\ee
where the notation $G (\cdots ]$ highlights that $G$ is a function of $x,y$, but a functional of $\phi_i$'s.
The
nonlinear generalizations of Onsager relations can then be written as 
\be \label{1con1}
G_{i j}^R (x, y; \phi_i (\vx)] = \eta_{\phi_i}\eta_{\phi_j} G_{j i}^R (\eta y, \eta x; \Theta\phi_i (\vx) ] ,
\ee
where $G_{i j}^R (x, y; \phi_i (\vx)]$ denotes $G_{ij}^R$ is the presence of time-independent of sources $\phi_k (\vx)$ and
$\Theta$ here should be understood as the extension of (\ref{tg1}) to time-independent field configurations.

\item Taking the external sources $\phi_r, \phi_a$ in the generating functional~\eqref{gen0} to be time-independent,
then to first order in $\phi_a$, the generating functional $W$ can be ``factorized,''
\begin{eqnarray}
&&W [\phi_r, \phi_a]\notag \\
&&=i  \int d^{d-1} \vx \, \vev{\sO_i (\om=0, \vx)}_\phi  \, \phi_{ai} (\vx) + \cdots \notag\\
 &&=
i \tilde W [\phi_1] - i \tilde W [\phi_2] + \cdots \ ,
\label{1fac}
\end{eqnarray}
where $\tilde W [\phi_i (\vx)]$ is some functional defined on the {\it spatial} part of the full spacetime, and satisfies
\be
\tilde W [\phi (\vx) ] = \tilde W [\Theta\phi (\vx)]  \ .
\ee
From~\eqref{1fac}, $\tilde W [\phi (\vx)]$ generates the zero-frequency limit of nonlinear response functions, thus it can be identified
as the partition function $\log Z$ of~\eqref{par0}.\footnote{Note that in $\tilde W$ the sources are still Lorentzian sources, so to relate to $\log Z$ one should also analytically these sources to Euclidean signature~\cite{Glorioso:2017lcn} (see also sec. IV B there for some examples).}

\een


\section{General structure of EFTs for local equilibrium} \label{sec:gen}

We now proceed to formulate the general structure of EFTs for observables defined on a CTP.
We restrict  our attention to  physical processes whose characteristic spacetime scales
are much larger than typical relaxation scales, i.e. local equilibrium systems.




\subsection{The non-equilibrium effective action} \label{sec:nonac}

Consider the generating functional~\eqref{gen0} expressed in terms of path integral of microscopic variables
\bln 
e^{W [\phi_1, \phi_2]}  
& =   \int_{\rho_0} D \psi_1 D \psi_2 \, e^{i I_0 [\psi_1, \phi_1] - i I_0 [\psi_2; \phi_2]}  \
\label{fpth}
\end{align}
where 
$\psi_{1,2}$  denote microscopic dynamical variables for the two segments of the CTP and $I_0 [\psi; \phi]$ the microscopic action
in the presence of external sources. The minus sign in the second term comes from the reversed time integration for the second segment. The trace in~\eqref{gen0} is implemented by imposing in the path integrals
\be \label{infbd}
\psi_1 (t_f) = \psi_2  (t_f) = \psi_f,  
\ee
at some $t_f  \to \infty$ and integrating over all values of $\psi_f$.

Now 
imagine separating the degrees of freedom in terms of UV and IR variables and integrating
out UV variables, we obtain an effective action of IR variables $\chi_{1,2}$ where now should come in two copies, one for each segment of the contour,
\be
e^{W [\phi_1, \phi_2]}  =  \int_{\ep} D \chi_1 D \chi_2 \, e^{i I_{\rm EFT} [\chi_1, \phi_1; \chi_2 , \phi_2; \rho_0]}
\label{left}
\ee
where $\ep$ is 
the UV cutoff of the effective theory.
In~\eqref{left}, we should also impose
\be \label{effbd}
\chi_1 (t_f) = \chi_2  (t_f)  = \chi_f, \quad t_f \to \infty \
\ee
and integrate over $\chi_f$.
As before it is convenient to introduce $r$-$a$ variables for $\chi_{1,2}$
\be
\chi_{r} = \ha (\chi_{1} + \chi_{2}), \qquad \chi_{a} = \chi_{1} - \chi_{2}
\ee
which we will take to be real.

 One can write down $I_{\rm EFT}$ as the most general  theory after specifying  appropriate dynamical variables  $\chi_{1,2}$, and symmetries/constraints to be satisfied. Compared with~\eqref{par1},
 the effective action for a non-equilibrium system in~\eqref{left} has some new features:


\ben

\item While the fundamental action in the path-integral in (\ref{fpth}) is factorized in terms of 1 and 2 variables (i.e. it has the form $I_0[\psi_1,\phi_1]-I_0[\psi_2,\phi_2]$), due to the initial condition from the state $\rho_0$ and the future boundary condition~\eqref{infbd},
integrating out the UV variables in general results in couplings between the IR variables $\chi_1$ and $\chi_2$, so that $I_{\rm EFT}[\chi_1,\chi_2]$ is no longer factorized. 


\item After integrating out the UV variables, the dependence on $\rho_0$ is also encoded in $I_{\rm EFT}$.
In particular, that the system is in local equilibrium should be reflected in the structure of $I_{\rm EFT}$.

\item There are also additional constraints due to unitary time evolution.  From
similar arguments which lead to the unitarity constrains~\eqref{top1},~\eqref{odd1} and~\eqref{poss} on the generating functions, we find that $I_{\rm EFT}$ should satisfy
\bega\notag
 I^*_{\rm EFT} [\chi_1 , \phi_1; \chi_2, \phi_2]\\ = - I_{\rm EFT} [\chi_2 , \phi_2; \chi_1, \phi_1]
 \label{fer1}\\
 \label{pos}
{\rm Im} \, I_{\rm EFT} \geq 0, \quad {\rm for \;\; any} \; \; \chi_{1,2} \\
\label{1keyp2}
I_{\rm EFT}[\chi_r, \phi_r; \chi_a = \phi_a =0 ]=0\ .
\end{gather}
The derivations for~\eqref{fer1}--\eqref{1keyp2} parallel those for~\eqref{top1},~\eqref{odd1} and~\eqref{poss} as one can treat
slow modes $\chi$'s as ``backgrounds'' for the fast modes.
 For more details on their derivation, see Appendix A of~\cite{GL}. 
Equation~\eqref{fer1} implies that terms in $ I_{\rm EFT}$ which are {\it even} under exchange of $1,2$ indices must be purely imaginary.\footnote{Note that the original factorized form of the action in~\eqref{fpth} is real and is odd under exchange of $1,2$ indices.} As one expects such even terms will generically be generated when integrating out fast variables, $I_{\rm EFT}$ is hence generically complex. Equation~\eqref{pos} then says the imaginary part of $I_{\rm EFT}$ is non-negative which ensures that
path integrals~\eqref{left} are well defined as $|e^{i I_{\rm EFT}}| \leq 1$. Finally equation~\eqref{1keyp2} implies that all terms in $I_{\rm EFT}$ must contain
at least one factor of $a$-fields.

\item Any symmetry of the fundamental action $I_0$ which is preserved by the state $\rho_0$ should also be imposed in the effective action $I_{\rm EFT}$. For a global symmetry whose transformation parameters are spacetime independent, the boundary condition~\eqref{effbd} implies $\chi_{1,2}$ must transform at the same time, i.e. there is only a {\it single} copy of global symmetry
in $I_{\rm EFT}$. For example if $I_0$ is parity invariant, then $I_{\rm EFT}$ should be invariant only under a simultaneous
parity transformation on $\chi_1$ and $\chi_2$. Note that~\eqref{effbd} does not constrain local transformations which vanish
at future infinity, so for local symmetries, $\chi_{1,2}$ should be able to transform independently, i.e. there can be two copies of them. \label{item:sy}

\item The only exception to the general statement of the last item is time reversal symmetry. Suppose both $I_0$ and $\rho_0$
are invariant under some time reversal transformation $\Th$.\footnote{As discussed before equation~\eqref{1newfdt1}  $\Th$ can be the time reversal $\sT$ itself, or any combinations of $\sC, \sP$ with $\sT$, such as $\sC\sP\sT$.}   $I_{\rm EFT}$ includes dissipative and retardation effects from integrating out fast modes which can be considered as a bath for slow variables. Thus in general $I_{\rm EFT}$ cannot be invariant under $\Th$. It turns out that the time reversal can be imposed along with the local equilibrium condition to which we will turn now.

\een

\subsection{Dynamical KMS symmetry}\label{sec:kms}

Now let us consider $\rho_0$ given by the thermal density matrix~\eqref{thro}.  The action $I_{\rm EFT}$ should be such that $W$ obtained from~\eqref{left} satisfies the condition~\eqref{1newfdt1} which encodes both time reversal symmetry $\Th$ and the KMS condition for thermal equilibrium.
This can be achieved by requiring $I_{\rm EFT}$ to satisfy an {\it anti-linear $Z_2$} symmetry, to which we will refer as the dynamical KMS symmetry (or local KMS symmetry).

The explicit form of the $Z_2$ dynamical KMS transformation of a slow variable depends on whether  it corresponds to a conserved quantity and whether there is a dynamical local temperature.  As an illustration let us consider here the simplest case. Consider a system
with $\rho_0$ given by a thermal density matrix with inverse temperature $\b_0$, and suppose the couplings between IR variables $\chi$ and the sources $\phi$ can be written in a linear form
\be\label{libc}
I_{\rm EFT} [\chi_1, \phi_1; \chi_2, \phi_2] =
\cdots + \int d^d x \, \le(\chi_1 \phi_1 - \chi_2 \phi_2 \ri)  
\ee
where $\cdots$ denotes terms in the action which depend on $\chi$'s and $\phi$'s separately. 
It can be readily checked by formal manipulations of path integrals\footnote{In reality one will encounter divergences which may spoil the validity of formal manipulations. One should make sure that there exists a regularization procedure which is compatible with the $Z_2$ symmetry.} that the generating functional~\eqref{left}  satisfies~\eqref{1newfdt1}  if we require that $I_{\rm EFT}$ satisfy
\be \label{lkms}
I_{\rm EFT} [\chi_1, \phi_1; \chi_2, \phi_2] =I_{\rm EFT}   [\tilde \chi_1, \tilde \phi_1; \tilde \chi_2, \tilde \phi_2]
\ee
where $\tilde \phi_{1,2}$ are given by~\eqref{tiV0} and
\be
\begin{split}
\tilde \chi_1 (x) &= \Th \chi_1 (t - i \th, \vx) , \\
\tilde  \chi_2 (x) &=  \Th \chi_2 (t + i (\beta_0 - \th), \vx ) \ .
\label{tiV1}
\end{split}
\ee
For the case of~\eqref{tiV} we then have
\be\begin{split} \label{tiV2}
\tilde \chi_1 (x) &= \chi_1 (-t + i \th, - \vx ) , \\
\tilde  \chi_2 (x) &=  \chi_2 (- t - i (\beta_0 - \th), -\vx )  \ .
\end{split}\ee
A symmetry like equation~\eqref{lkms} is sometimes called a spurious symmetry as one relates the action for one set of sources to another. In the absence of external sources~\eqref{tiV1} becomes a genuine $Z_2$ symmetry of the action.
The quantum form~\eqref{tiV1} as a symmetry to impose thermal equilibrium was recently also advocated in~\cite{Sieberer2}.

The simple form of the coupling~\eqref{libc} applies only to non-conserved IR variables, such as order parameters near a critical points. For $\chi$'s associated with conserved quantities, as we will see in later sections,  the couplings are more complicated, especially in cases with local dynamical temperature.
The corresponding forms of dynamical KMS symmetry are also more intricate~\cite{CGL,CGL1}.
We will discuss in detail the explicit forms of the dynamical KMS transformations for those cases in Sec.~\ref{sec:eft1}--Sec.~\ref{sec:hydroeft}, and work out explicitly the constraints the dynamical KMS symmetry imposes on $I_{\rm EFT}$.


 One remarkable consequence of the $Z_2$ dynamical KMS symmetry--which does not depend on the specific form of the transformations and class of theories, is that when combined with unitarity constraints~\eqref{fer1}--\eqref{1keyp2} it implies the existence of an ``emergent'' entropy current whose divergence is non-negative~\cite{GL}.
We will review this story in Sec.~\ref{sec:entropy}.





\subsection{Example: Brownian motion}\label{sec:browni}

Before proceeding with a further discussion of the general structure of $I_{\rm EFT}$, let us consider a simple
example.

Consider a free particle placed in contact with a bath of harmonic oscillators~\cite{CLmodel}. The microscopic action of the system is $I_0=\int dt \, L_0$, with
\be\label{jnne}
 L_0={M \ov 2} \dot x^2+\sum_{i=1}^n\ {m_i \ov 2} (\dot q_i^2-\omega_i^2 q_i^2)+  x\sum_{i=1}^n\lambda_i  q_i\ \ee
where dot denotes time derivative and the sum $\sum_i$ should be understood as an integral in the case that the bath has a continuous spectrum. The CTP path integral for the system is then given by
\be \label{sgjn}
Z = \int Dx_1 Dx_2 \int_{\rho_0} \prod_{i=1}^n (D q_{1i} D q_{2i} ) \; e^{iI_0[x_1,q_1]-i I_0[x_2,q_2]}
\ee
where $\rho_0$ is the thermal density matrix with inverse temperature $\b_0$.
Now let us assume $M$ is big such that the motion of the particle is much slower than those of the bath of oscillators, and
we can then integrate out the bath to obtain
\be \label{seffp}
Z =\int  Dx_1 Dx_2 \, e^{i I_{\text{EFT}}[x_1,x_2]}\ ,
\ee
where $I_{\text{EFT}}[x_1,x_2]$ is the resulting effective action for slow modes $x_1, x_2$.

Since $I_0$ is quadratic, the integrals for $q_i$ can be performed explicitly.\footnote{See Sec. 3.2 of \cite{Kamenev} for details on the exact evaluation.} 
Here we will deduce the form of $I_{\text{EFT}}[x_1,x_2]$  as an EFT from symmetries: (i) Since the $I_0$ is quadratic, $I_{\rm EFT}$ should also at most be quadratic in $x_r = \ha (x_1 + x_2)$ and $x_a = x_1 - x_2$; (ii) 
 Since $I_0$ is invariant under $x \to -x, q_i \to - q_i$, $I_{\rm EFT}$ should be invariant under $x_{r,a} \to - x_{r,a}$ (recall item~\ref{item:sy} of Sec.~\ref{sec:nonac}); (iii) for large $M$ we should be able to expand $I_{\rm EFT}$ in time derivatives.
With these considerations, the most general $I_{\text{EFT}}=\int dt \, L_{\text{EFT}}$ satisfying \eqref{fer1} and \eqref{1keyp2} can be written as
\be \label{eftbr}
L_{\text{EFT}} = - c x_a x_r -\nu x_a \dot x_r+ \tilde M \dot x_r \dot x_a+\frac i2 \sigma x_a^2+\cdots
\ee
where $\cdots$ denote terms with higher derivatives. Furthermore~\eqref{pos} requires that
\be \label{inm}
\sig \geq 0 \ .
\ee
Finally we should impose the dynamical KMS symmetry. Applying~\eqref{tiV1} to $x_{1,2}$ and expanding it to leading order in derivatives we find that
\be\begin{split} \label{xrn}
\tilde x_{r} (-t)  &= x_{r }(t) + \cdots , \\ \tilde x_{a } (- t) &= x_{a} (t) + i \beta_0   \p_t x_r (t) + \cdots \ .
\end{split}\ee
 Requiring~\eqref{eftbr} to be invariant under~\eqref{xrn} leads to
\be \label{propI}
\nu=\frac 12\sigma \beta_0 \geq 0
\ee
where in the second inequality we have used~\eqref{inm}.

To see the connection of~\eqref{eftbr} with the standard description in terms of Langevin equation, let us consider a Legendre transformation with respect to $x_a$, i.e. write
\be\label{ll1}
 \frac i2 \sigma x_a^2=\frac i2 \frac 1\sigma \xi^2+\xi x_a,\ee
where $\xi$ is a new dynamical variable, and the path integral (\ref{seffp}) becomes
\be \int Dx_a Dx_r D\xi \, e^{i\int dt\left(- c x_a x_r  -\nu x_a \dot x_r+ \tilde M \dot x_r \dot x_a+\frac i2 \frac 1\sigma \xi^2+\xi x_a\right)}\ .\ee
The dependence on $x_a$ in the exponent is now linear, i.e. it is a Lagrange multiplier.
Integrating it out reduces the path integral to
\be \int D\xi  Dx_r \, \delta(\tilde M \ddot x_r+ \nu \dot x_r + c x_r -\xi)\,  e^{-\int dt\frac 1{2\sigma}\xi^2}\ ,\ee
which is equivalent to the Langvin equation
\be
\tilde M \ddot x_r+\nu \dot x_r + c x_r=\xi\ ,
\ee
with a stochastic force $\xi$ which is Gaussian distributed with variance $\sigma$,
\be \label{ll2}
\langle \xi(t)\xi(0)\rangle = \sigma \delta(t)\ .
\ee
Note that:

\begin{itemize}

\item $x_r$ corresponds to physical position of the particle, while $x_a$ is the Legendre conjugate of the stochastic force and thus should be interpreted as position ``noise''.

\item  $\tilde M$ can be interpreted as the effective mass of the particle (which is in principle renormalized from its ``bare'' value $M$ due to interactions with the bath). $c x_a x_r$ term
is an induced potential from coupling to the bath (notice that the coupling~\eqref{jnne} is not invariant under a translation of $x$).

\item $\nu$ is the friction coefficient, i.e. in general there are dissipative terms.

\item The imaginary part of the action~\eqref{eftbr} describes fluctuations, with ${1 \over \sig}$ controls the scale of fluctuations of position noise $x_a$ (equivalently $\sigma$ controls the magnitude of fluctuations of the stochastic force).

\item The relation~\eqref{propI} is precisely the Einstein relation which relates the dissipative coefficient to the variance of the fluctuating force, here arising as a consequence of the dynamical KMS symmetry.

\item Non-negativity of $\text{Im}\,I_{\text{EFT}}$ together with the dynamical KMS symmetry implies non-negativity of $\nu$, which in turn guarantees causality.\footnote{Here causality simply means responses come after disturbances.}



\end{itemize}
Finally we should mention that  there is no $\nu$ or $\sig$ term generated if the number $n$ of bath oscillators is finite.
There is a nonzero $\sig$ and $\om$ only when there is a continuous spectrum of oscillators with frequencies starting from $\om =0$, i.e.  to generate dissipations one needs a continuum of low frequency modes. Physically this makes sense: only when there is a continuum of modes, can energy disappear without a trace, thus having genuine dissipations.  Note that when there are nonlinear interactions among bath degrees of freedom or nonlinear couplings between the heavy particle and the bath, in general nonlinear terms will be induced and $\cdots$  in~\eqref{eftbr} could have cubic and higher order terms in $x_{r,a}$.



\subsection{General structure of equations of motion}
\label{sec:fluc}

The Brownian motion example we discussed in last section is very simple,
but the structure is completely general. Now we show that a general theory~\eqref{left}
has a parallel structure. In particular, one can identify $\chi_r$ as representing standard physical quantities, while $\chi_a$ the stochastic counterparts (noises). For convenience let us set the sources to zero.

Equation~\eqref{1keyp2} implies that all terms in the action should contain at least one factor of $\chi_a$.
To make this manifest,  we can expand  the Lagrangian in $\chi_a$ as
\be \label{effla}
\sL_{\rm EFT} = E [\chi_r] \chi_a + {i \ov 2} \chi_a \hat F[\chi_r, \p] \chi_a + O(\chi_a^3)
\ee
where $E [\chi_r]$ is a local function of $\chi_r$ and their spacetime derivatives, and similarly
for $\hat F[\chi_r, \p]$ except that it can also contain derivatives acting on the second factor of $\chi_a$, i.e. $\hat F$ is a local differential operator.  By definition, $\hat F$ should be a non-negative operator following~\eqref{pos}, and can be taken to
satisfy (up to total derivatives)
\be
\hat F [ \chi_r, \p]  = \hat F^* [ \chi_r, \p]
\ee
where $\hat F^*$ denotes the differential operator obtained by shifting all derivatives from the second $\chi_a$ to the first $\chi_a$ in~\eqref{effla}. From~\eqref{fer1}
the first term in~\eqref{effla} is real while the second term is pure imaginary.
Now equations of motion from variations of $\chi_a$ can be written as
\be
E [\chi_r]  + O(\chi_a) = 0
\ee
while those from variations of $\chi_r$ will contain at least one factors of $\chi_a$ for all terms.
Considering also the boundary condition~\eqref{effbd}, we can thus consistently  set
\be \label{nios}
\chi_a = 0 \ .
\ee
In this case we will then have $\chi_1 = \chi_2 = \chi_r \equiv \chi$ and equations of motion reduce to
\be \label{feq}
E[\chi] = 0 \ .
\ee
In other words, in equations of motion there is only one copy of dynamical variables and thus $\chi_r$ can be identified
as representing standard physical quantities.

Following similar steps as~\eqref{ll1}--\eqref{ll2} one again finds that to quadratic order in $\chi_a$ expansion,~\eqref{left}
is equivalent to a stochastic equation with a multiplicative noise\footnote{Since $\hat F$ is a non-negative operator, it is well-defined to take its ``square root.''}
\be \label{jje}
E[\phi, \chi] = \hat F^\ha  [\chi, \p] \xi, \qquad \vev{\xi (x) \xi (0)} = 
\de^{(d)} (x)  \ .
\ee

Stochastic equations such as~\eqref{jje} are usually written down phenomenologically. Except for near equilibrium situations, the structure of multiplicative factor $\hat F^\ha$ could only be deduced by guesswork. The EFT formalism provides a systematic way to derive such factors for far-from-equilibrium situations. Furthermore, the action~\eqref{effla} provides
full non-Gaussian structure for noises in terms of $O(\chi_a^3)$ and higher, which cannot be captured using~\eqref{jje}.

In the MSR formalism~\cite{msr,Dedo,janssen1}, a path integral is obtained by introducing a Lagrange multiplier to
exponentiate the stochastic equation (\ref{jje}), and thus the physical content of the resulting theory is exactly the same as~\eqref{jje}.\footnote{See \cite{Kamenev} for a nice introduction of the MSR formalism, and its relation with the Schwinger-Keldysh formalism.}

The above discussion completely goes through with {\it physical sources}  turned on, which have $\phi_1 =\phi_2$~(i.e. $\phi_a =0$).
Turning on $\phi_a$ will source the equations for $\chi_a$ and thus $\chi_a$ will no longer be zero. Thus $\phi_a$ can be interpreted as stochastic sources.

\subsection{Classical limits} \label{sec:hbar}


So far our discussion is at full quantum level with a finite $\hbar$. The path integrals~\eqref{left} include both
statistical and quantum fluctuations.  In many situations, say at a sufficiently high temperature, quantum fluctuations may be neglected. One could then restrict to
the classical limit $\hbar \to 0$.

Before discussing how to take the classical limit in the effective action $I_{\rm EFT}$, let us first consider
how to take the classical limit 
in the full generating functional~\eqref{gen0}. Here are the basic inputs:

\ben

\item Restoring $\hbar$, the exponent on right hand side of~\eqref{gen0} should have an overall factor of ${1 \ov \hbar}$.

\item Expanding $W$ in powers of $\phi_r, \phi_a$, schematically (we suppress all spacetime integrations and index summations, etc.)
\be \label{pmm}
W = \sum_{m,n=0}^\infty G_{r^m a^n} \phi_a^m \phi_r^n
\ee
where $G_{r^m a^n}$ denotes a Green function with $n$ $a$-subscripts and $m$ $r$-subscripts and has the schematic form
\be
G_{r^m a^n} = {1 \ov \hbar^{m+n}}  \vev{\sO_r^m \sO_a^n}  \ .
\ee
As discussed below~\eqref{yepi}, when written in terms of standard operator orderings, $G_{r^m a^n}$ contains
$n$ commutators.
In the classical limit each commutator becomes $\hbar$ times the corresponding Poisson bracket, and as a result
$G_{r^m a^n} $ scales with $\hbar$ as
\be
G_{r^m a^n} \sim {1 \ov \hbar^{m+n}}  \hbar^n \sim {1 \ov \hbar^m}  \ .
\ee

\een
Thus in order for~\eqref{pmm} to have a sensible classical limit we need to scale
\be
\phi_r \to \phi_r , \qquad \phi_a \to \hbar \phi_a  \ .
\ee
i.e. we take the external sources to be
\be \label{lim1}
\phi_1 = \phi + {\hbar \ov 2} \phi_a , \qquad \phi_2 = \phi - {\hbar \ov 2} \phi_a\ ,
\ee
as $\hbar\to 0$, where $\phi,\phi_a$ are finite. This makes sense as $\phi_r$ correspond to physical sources we turn on and thus should not scale in the classical limit.

Now the slow dynamical variables $\chi$'s must have the same scaling
\be \label{lim2}
\chi_1 = \chi + {\hbar \ov 2} \chi_a , \qquad \chi_2 = \chi - {\hbar \ov 2} \chi_a\ ,
\ee
as $\hbar\to 0$, where $\phi,\phi_a$ are finite. There are two ways to argue for this. The first is that slow variables themselves can be viewed as ``external sources'' for fast variables. The second is that presence of $\phi$'s induces $\chi$'s, e.g. through couplings like~\eqref{libc}, and thus they should have the same $\hbar$ scaling.
Similarly with $I_{\rm EFT}$ written schematically as
\be
{1 \ov \hbar} I_{\rm EFT} = \sum_{m,n} g_{mn}  \Lam_r^n \Lam_a^m
\ee
where $\Lam_{a,r}$ denote collectively both sources and dynamical variables, then $g_{mn}$ should have the same ``semi-classical'' expansion as  $G_{r^m r^n}$ in~\eqref{pmm}, i.e.
\be
g_{mn} \sim {1 \ov \hbar^m} \le(1 + O(\hbar) + O(\hbar^2) + \cdots \ri), \qquad \hbar \to 0  \ .
\ee
Thus in the limit~\eqref{lim1}--\eqref{lim2}, ${1 \ov \hbar} I_{\rm EFT}$ should have a finite limit, and the path integral~\eqref{left} survive the $\hbar \to 0$ limit.

That the path integral~\eqref{left} should survive in the $\hbar \to 0$ limit should come as no surprise  as the system still has statistical fluctuations. This is also familiar in the equilibrium context in the Euclidean path integral~\eqref{par0} for a partition function with the Euclidean action $I_0$ having the form
\be
{1 \ov \hbar} I_0 = {1 \ov \hbar} \int_0^{\hbar \beta} d\tau \int d^{d-1} x  \, \sL_0 
\ee
In the $\hbar \to 0$ limit with $\b$ fixed, the range of Euclidean time $\tau$ goes to zero and to lowest order we can take all fields in $\sL_0$ to be independent of  $\tau$, i.e.
\be \begin{gathered}
{1 \ov \hbar} I_0 =  {1 \ov \hbar} \hbar \beta  \int d^{d-1} x  \, \sL_0
= \beta I_{\rm classical}  , \quad \hbar \to 0, \\
I_{\rm classical} =  \int d^{d-1} x  \, \sL_0  \ .
\end{gathered}\ee
In the classical limit, there is some effective $\hbar_{\rm EFT}$ which controls the loop expansion of the path integral~\eqref{left}.
Again the equilibrium situation should yield a hint: in equilibrium the statistical fluctuations are controlled by
\be
\heff \propto {1 \ov \sN}
\ee
with $\sN$ is the number of degrees of freedom.

Finally note that since equations of motion ignore all fluctuations including both quantum and statistical ones,
thus they should be interpreted as describing the thermodynamic limit.



\subsection{Dynamical KMS symmetry in the classical limit}\label{sec:dkms}

In this subsection we elaborate on the general structure of the EFT action under the dynamical KMS symmetry in the class limit. Such a structure plays an important role in many subsequent discussions.

In the classical limit, the dynamical KMS transformations has a particularly
simple structure. As an example, consider ~\eqref{tiV0} and (\ref{tiV1}) in the classical limit~\eqref{lim1}--\eqref{lim2} for which we should also restore $\hbar$ in $\b_0$ as $\hbar \b_0$. Taking $\hbar \to 0$ with $\b_0$ finite we then find that\footnote{Note
$\Th \p_0 \phi (x) = (\p_0 \phi) (-x)$.}
\bega
 \tilde \phi_r(x)=\Theta \phi_r( x),\; \tilde \phi_a(x)=\Theta \phi_a(x)+i\Theta\beta_0\p_0\phi_r(x) \\
 \label{hns}
 \tilde \chi_r(x)=\Theta \chi_r( x),\; \tilde \chi_a(x)=\Theta \chi_a(x)+i\Theta\beta_0\p_0\chi_r(x)\ ,
 \end{gather}
where $\Th$ is the anti-linear time-reversal transformation described earlier. Note that the above equations are exact in the limit $\hbar\to 0$: it is a finite transformation and we did not perform derivative expansion.


More generally, as will be discussed explicitly in later sections, the dynamical KMS transformations have the following structure in the classical limit
\be \label{dkms}
\tilde \Lam_r = \Th   \Lam_r , \qquad  \tilde \Lam_a = \Th  \Lam_a + i \Th \Phi_r
\ee
where again $\Lam_{r,a}$ denote both dynamical variables and sources, and
$\Phi_r$ denotes some expression of $r$-variables
which transforms under $\Th$ the same way as the corresponding $\Lam$ except with an additional minus sign, and {\it contains a single derivative}.
It can be readily checked with these properties and $\Th^2 = 1$, the transformation is $Z_2$,
\be
\tilde{\tilde \Lam}_a = \Th \tilde \Lam_a + i \Th \tilde \Phi_r =
\Th^2 \Lam_a + i \Th^2 \Phi_r  - i \Th^2 \Phi_r = \Lam_a  \ .
\ee


An important feature of~\eqref{dkms} is that it preserves the sum of the numbers of $a$-indices and derivatives.
This motivates us to introduce the expansion
\be
\label{Ll}\mathcal L_{\rm EFT}=\sum_{l=1}^\infty \mathcal L_l,\qquad \mathcal L_l \equiv \sum_{n+m=l}\mathcal L^{(n,m)}\ ,
\ee
where $\mathcal L^{(n,m)}$ contains precisely $n$ factors of $a$-variables and $m$ spacetime derivatives. For example, $\mathcal L_1=\mathcal L^{(1,0)}$ contains one factor of $a$-variables with no derivatives, $\mathcal L_2=\mathcal L^{(2,0)}+\mathcal L^{(1,1)}$ contains either two facotors of $a$-variables with no derivatives, or one factor of $a$-variables with one derivative, and so on.\footnote{As noted below equation~\eqref{jje}, the MSR formalism for stochastic equations can only treat Gaussian noises and thus does not capture $\mathcal L_l$ for $l>2$.} 

Under~\eqref{dkms}, $\sL_l$ transforms to itself, thus each $\sL_l$ must be separately invariant. That is, under the dynamical KMS transformation it must change by a total derivative
\be
\tilde \sL_l - \sL_l = \p_\mu W^\mu_l, \quad \qquad  \tilde{\mathcal L}_l \equiv \mathcal L_l [\Theta\tilde \Lambda_a,\Theta\tilde\Lambda_r] \ .
\ee
In examples of later sections we will always organize $\sL_{\rm EFT}$ in terms of~\eqref{Ll}.

The $Z_2$ structure of the dynamical KMS symmetry also implies the following important structure.
Consider a Lagrangian $\sL$ which satisfies all the unitarity constraints~\eqref{fer1}--\eqref{1keyp2} and we want to impose the dynamical KMS symmetry.
 Due to the $Z_2$ nature of the dynamical KMS transformation, then
\be \label{LLc}
\mathcal L_{\rm EFT}=\frac 12 \left(\mathcal L+\tilde{\mathcal L} \right), \qquad  \tilde{\mathcal L} \equiv \mathcal L [\Theta\tilde \Lambda_a,\Theta\tilde\Lambda_r]
\ee
automatically satisfies dynamical KMS invariance.  Note, however, that $\tilde{\mathcal L}$ contains terms with $r$-fields only
due to the $- i \Th \Phi_r$ term in $\tilde \Lam_a$.   Thus constructed $\mathcal L_{\rm EFT}$ violates the condition (\ref{1keyp2}). We must then further require that pure $r$-terms in $\tilde{\mathcal L}$ must be a total derivative,
\be\label{v00} 
\tilde \sL \bigr|_{\Lam_a =0}  = i \p_\mu V^\mu_0\ ,
\ee
where $V^\mu_0$ is some vector (it is real) which does not contain any $a$-fields.
The form (\ref{LLc}) together with (\ref{v00}) is enough to ensure $\sL_{\rm EFT}$ is invariant under both unitarity constraints and dynamical KMS.

To conclude this discussion we mention by passing that interestingly one can show equation~\eqref{v00} is equivalent to the condition that $\sL$ is supersymmetrizable~\cite{ping}.

\subsection{Second law, emergent entropy and non-dissipative theories} \label{sec:entropy}

The combination of unitarity constraints~\eqref{fer1}--\eqref{1keyp2} and dynamical KMS symmetry leads to a remarkable
consequence: there exists an ``emergent'' entropy which satisfies the second law, through a Noether-like procedure~\cite{GL}. 
The result only depends on the general structure exhibited in~\eqref{dkms}, not the the specific form of the dynamical KMS transformations.  We will review the main results whose derivation we will refer readers to~\cite{GL}.

Invariance under dynamical KMS symmetry
\be \label{lkms11}
I_{\rm EFT} [\Lam_r, \Lam_a] =I_{\rm EFT}   [\tilde \Lam_r, \tilde \Lam_a]
\ee
implies that the corresponding Lagrangian density should change as a total derivative
\be\label{344}
\tilde \sL_{\rm EFT} = \sL_{\rm EFT}  + \p_\mu V^\mu
\ee
where $\tilde{\mathcal L}_{\rm EFT}=\mathcal L_{\rm EFT} [\Theta\tilde \Lambda_a,\Theta\tilde\Lambda_r]$.
$V^\mu$ can be expanded in the number of $a$-fields as
\be\label{345}
V^\mu = i V_0^\mu+V_1^\mu+\cdots \
\ee
where $V_k^\mu$ contains $k$ factors of $a$-fields. Now with the Lagrangian written in the form of  (\ref{LLc}) and satisfying~\eqref{v00},
one has  $V_k^\mu=0$ for $k>0$, i.e. only $V_0^\mu$ survives and given by~\eqref{v00}. However, it is often convenient to use integration by parts to write
terms linear in $\chi_a$ as in the first term of~\eqref{effla} with no derivatives acting on $\chi_a$. This may generate a nonzero $V_1^\mu$.
Thus it is possible to write the Lagrangian density of the form~\eqref{effla}
and to have all the $V_k^\mu =0$ for $k > 1$.

 For such a Lagrangian consider the current
\be \label{endn}
s^\mu = V_0^\mu - \hat V_1^\mu
\ee
where $\hat V_1^\mu$ is obtained from $V_1^\mu$ by replacing all the $\Lam_a$ by $\Phi_r$.
Then one can show upon using equations of motion~\eqref{feq}
\be \label{lo2n}
\p_\mu s^\mu =\ha  \Phi_r \hat F (\chi_r, \p) \Phi_r  + \cdots   \
\ee
where $\cdots$ denote terms which depend on coefficients of terms of order $\chi_a^3$ and higher in~\eqref{effla}.
One can show by using~\eqref{pos} that the right hand side of~\eqref{lo2n} is non-negative order by order in derivative expansion.

Furthermore, it is in fact possible to resum all terms on the right hand side of~\eqref{lo2n} to all derivative orders, and
by using $Z_2$ dynamical KMS symmetry over and over,
to show
\be \label{n2n}
\De S \equiv  \int_{t=t_f} d^{d-1}x \, s^0-\int_{t=t_i} d^{d-1}x \, s^0  = \sR \geq 0\ ,
 \ee
where $t_f>t_i$, and $\sR$ is an integral transform of ${\rm Im} \, I_{\rm EFT}$ which preserves its non-negativity~\eqref{pos}:
\be\begin{split} \mathcal R\equiv &\int dz \frac{\pi}{\sinh^2(\pi z)}\times\\
&\times\left[\frac 12(\cosh(\pi z)-1)(F(z)+F_\Theta(z))+F(z)\right]\ ,\end{split}\ee
and
\be \begin{split}F(z)&\equiv \text{Im}\, I_{\text{EFT}}[\Lambda_r,z\Phi_r],\\ F_\Theta(z)&\equiv \text{Im}\, I_{\text{EFT}}[\Theta\Lambda_r,z\Theta\Phi_r]\ .\end{split}\ee
The non-negativity of $\mathcal R$ clearly follows from (\ref{pos}). The conclusion~\eqref{n2n} thus holds non-perturbatively in derivatives.



This result  implies that there exists a monotonically increasing quantity with time. In the next few sections we will apply the procedure to various classes of theories and show that the quantity coincides with the standard thermodynamic entropy in the equilibrium limit.

In the traditional formulation of hydrodynamics, as reviewed in Appendix \ref{app:hydro}, the local second law $\p_\mu s^\mu\geq 0$ is imposed as a phenomenological constraint. Here we showed that the second law follows from  fundamental properties of a quantum system. (See also~\cite{Jensen:2018hse} for a recent discussion of the condition~\eqref{pos} and constraints from the second law.)


With an off-shell definition of the entropy current~\eqref{endn} and its divergence~\eqref{lo2n}, we can define
a non-dissipative theory as one which satisfies $\p_\mu s^\mu =0$. In other words, for such a theory all the
coefficients of the Lagrangian which contribute to the right hand side of~\eqref{lo2n} have to vanish.
In the Brownian motion example of Sec.~\ref{sec:browni} this corresponds to
$\nu$ and $\sig$ being zero. More generally, from~\eqref{lo2n} one sees that this non-dissipative condition is
equivalent to the statement that the Lagrangian can be written as
\be \label{nemp}
\sL_{\rm EFT} = E_{\rm NP} [\chi_r] \chi_a
\ee
which is {\it invariant under dynamical KMS symmetry by itself}. In such a Lagrangian there are no terms which are quadratic order in $\chi_a$ and higher. In contrast, for a general Lagrangian~\eqref{effla}, dynamical KMS symmetry relates certain coefficients in $E[\chi_r]$ with those in $\hat F[\chi_r, \p]$ and higher order terms of $\chi_a$. It is precisely those coefficients in $E[\chi_r]$
which are dissipative. In the absence of $O(\chi_a^2)$ and higher order terms  one could also see that the dynamical KMS transformation essentially enhances to a continuous symmetry. The conservation of $s^\mu$ can then be understood from the standard Noether procedure.

Given the structure~\eqref{nemp} it is tempting to conjecture that for a non-dissipative theory one could factorize the Lagrangian~\eqref{nemp} in a form
\be \label{nemp1}
\sL_{\rm EFT} =  E_{\rm NP} [\chi_r] \chi_a  = \sL_s [\chi_1] - \sL_s[\chi_2] + O(\chi_a^3)
\ee
for some local Lagrangian $\sL_s$.  We will present support for such a factorization in various examples in later sections, including ideal fluids.  But we note that this statement appears to be not true for systems with anomalies. It can be shown that
the action in~\cite{Glorioso:2017lcn} cannot be factorized even in the non-dissipative limit.

\subsection{Role of ghosts and supersymmetry}

While it can be readily seen that~\eqref{1keyp2} leads to~\eqref{top1} at tree-level of the path integral,
in~\cite{CGL} it was realized that loop corrections could potentially violate~\eqref{top1}.
Anticommuting ghost variables and BRST symmetry were then introduced to make sure the unitarity constraint~\eqref{top1} is maintained to all loops~\cite{CGL} (see also~\cite  {Haehl:2015foa,Haehl:2015uoc,yarom}).
Intriguingly, it can be further shown that when the BRST symmetry is combined with the condition~\eqref{v00} from the $Z_2$ dynamical KMS symmetry, there is always an emergent supersymmetry~\cite{CGL,ping} and, on the converse,  supersymmetry can be used to impose~\eqref{v00}~\cite{ping,yarom}.\footnote{See also~\cite{Haehl:2015foa,Haehl:2015uoc} which used supersymmetry as an input for constructing an action principle for hydrodynamics. But note that since supersymmetry only imposes~\eqref{v00} it is not enough to ensure the full dynamical KMS symmetry.}
The presence of BRST and supersymmetries in EFTs can be considered as natural extensions of their appearance in earlier work
on functional approach to stochastic systems and others ~\cite{parisi,feigelman,Gozzi:1983rk,Mallick:2010su,zinnjustin,gozzib,gozzi1,gozzi2,camille1,camille2}, where connections between supersymmetries and fluctuation-dissipation
relations as well as Onsager relations have been long recognized.

In the context of stochastic systems, the introduction of ghost variables in the functional approach is not needed when one uses the Ito procedure to discretize the stochastic equation and thus the path integral.\footnote{There are further ambiguities when performing field redefinitions in the path-integral, see \cite{Kamenev} for a detailed discussion.} The procedure ensures that the Jacobian for  exponentiating a stochastic equation is upper triangular and thus has unit determinant.  For an EFT defined in the continuum, such a procedure cannot be used.
Recently a new regularization procedure was introduced in~\cite{Gao:2018bxz} and it was shown to all loop orders that: (i) Even in the absence of ghost variables, unitarity is maintained; 
(ii) Integrating out the ghost action results in no contributions. 
Thus ghost variables can be neglected. One key element in the discussion of~\cite{Gao:2018bxz} is the retarded nature of certain class of propagators in an EFT.\footnote{Note that the retarded structure of the propagator causes ghost diagrams to vanish is well-understood in the context of the Langevin equation, see e.g.~\cite{Arnold,Gonzalez}. Also the importance of retarded nature of propagators  in perturbation theory for a microscopic theory defined on a CTP was also well known, see e.g.~\cite{Kamenev}.}

In the formulation of~\cite{GL,CGL,CGL1}, the retarded nature of the propagators 
is a consequence of the $Z_2$ dynamical KMS symmetry and unitarity constraints,
and reflects the coincidence of thermodynamic and causal arrow time~\cite{GL}.
More explicitly,  it means that dissipative coefficients of the action must have the ``right'' signs--for example, friction coefficients, viscosities, conductivities must be non-negative--which ensures that on the one hand entropy increases monotonically with time,
and on the other hand the system is causal. We will see these features in explicit examples in the following sections.

Ghosts and supersymmetry could still be useful if one prefers to use other types of regulators which break the retarded structure of various propagators or dynamical KMS symmetry. They will help to ensure the normalization condition and part of the dynamical KMS symmetry to be manifestly preserved.  In this manner, they are pure book-keeping devices, not playing any role in low energy dynamics. In \cite{Jensen:2018hse} (see also \cite{Jensen:2018hhx}), supersymmetry was used to obtain a ``supercurrent'' version of the entropy current. Such supercurrent is conserved, in the sense that its superspace divergence vanishes.

\subsection{Organization of examples}

We now proceed to discuss explicit formulations of EFTs for various quantum statistical systems.
We will consider systems with spacetime translational and rotational symmetries, i.e. in a liquid phase. Energy and momentum are always conserved. Energy density disturbances lead to local temperature variances, i.e. a dynamical temperature.
We will organize our discussion into three class of systems (recall the discussion of slow variables in Sec.~\ref{sec:1b}):

 \ben

 \item With fixed background temperature and no conserved quantities. In this class of examples, one considers a system near a (finite temperature) critical point whose order parameters do not involve conserved quantities. If the fluctuations of conserved quantities
 are neglected, the macroscopic dynamics of the system is then controlled solely by that of the order parameters at a constant temperature. As an example in Sec.~\ref{sec:eft1} we will discuss model A of~\cite{hohenberg}, which describes an $O(n)$ vector model near
 its critical point. The corresponding non-equilibrium EFT can be used to study dynamical critical phenomena~\cite{hohenberg,Folk}.

\item With conserved quantities and a fixed background temperature. Consider, for example, a system with a $U(1)$  global symmetry, and we are interested in transports associated with conserved $U(1)$ charge. When one neglects effects from energy and momentum disturbances, the macroscopic dynamics of the system is then captured by that for the $U(1)$ current at a fixed background temperature, which we will examine  in Sec.~\ref{sec:eft2}. This approximation is sometimes referred to as the probe limit. The probe limit works well, for example, for a system with particle-hole symmetry, in which case charge flow and momentum flow are decoupled.


\item With conserved quantities and dynamical temperature. One now considers the full energy-momentum disturbances.
As an illustration in Sec.~\ref{sec:hydroeft} we will consider the resulting fluctuating hydrodynamics assuming the only conserved quantities are the stress tensor for a relativistic system.

\een
With these representative examples, it is straightforward to combine the elements of  Sec.~\ref{sec:eft1}--\ref{sec:hydroeft} to  general situations involving multiple conserved quantities and also non-conserved order parameters. For technical details regarding the formulation of a charged fluid the reader should consult the original papers \cite{CGL,CGL1}.



\section{EFTs I: critical $O(n)$ model}  \label{sec:eft1}


In this section we consider the critical dynamics of a $n$-component real order parameter $\chi_\rmi, \rmi=1,\cdots , n$ (i.e. model A~\cite{hohenberg,Folk}) at a fixed inverse temperature $\beta_0$.
As an illustration we will consider the classical limit and $\Th = \sC \sP \sT$ for which the dynamical KMS transformation~\eqref{tiV1} simplifies to
\be\begin{split} \label{ckms1}
\tilde \chi_{r \rmi} (x)  &= \chi_{r  \rmi}(-x) , \\
\tilde \chi_{a \rmi} (- x)& = \chi_{a \rmi} (x) +{ i \beta_0  } \p_0 \chi_{r \rmi} (x) \ . 
\end{split}\ee
The discussion below follows that of Appendix D of~\cite{GL} where readers may find more details.

Since~\eqref{ckms1} only involves time derivative we can treat time and spatial derivatives separately.
The Lagrangian can be then expanded in powers of $a$-fields and {\it time} derivatives as in~\eqref{Ll}
\be
\sL_{\rm EFT} = \sL_1 + \sL_2 + \cdots
\ee
where the subscripts count the total number of $a$-fields and time derivatives.
More explicitly, the most general form for $\sL_1$ can be written as
\be \label{ii0}
\sL_1 = E^\rmi_0 \chi_{a \rmi}
\ee
where $E^\rmi_0$ is a local functions of $\chi_{r \rmi}$ and their spatial derivatives (but no time derivative). Invariance under~\eqref{ckms1} then requires that there exists a local function $F$ from which
$E^\rmi_0$ can be obtained as (in~\eqref{ii1} there are only spatial integrations)
\be\begin{split} \label{ii1}
E_0^\rmi &= - {\de \sF \ov \de \chi_{r \rmi}}, \\
\sF (t; \chi_r]& = \int d^{d-1} \vec x \, F (\chi_r (x), \p_i \chi_r (x), \cdots) \ .
\end{split}\ee
In the first equation of~\eqref{ii1}, the minus sign is for later convenience. For $\sL_2$, after imposing the dynamical KMS symmetry, one finds at zeroth order for spatial derivatives
\be \label{ii2}
\sL_2 = -  \beta_0 f^{\rmi \rmj}  \p_0 \chi_{r \rmj}  \chi_{a \rmi}
+ {i } f^{\rmi \rmj} (\chi_r)   \chi_{a \rmi}   \chi_{a \rmj } + \cdots
\ee
where $\cdots$ denotes terms with higher spatial derivatives and $f^{\rmi \rmj} = f^{\rmj \rmi} $ are functions of $\chi_r$.
Note that the dynamical KMS symmetry relates the coefficients of the first term which is dissipative (as it contains only one time derivative) with that of the second term which controls fluctuations of $\chi_{a \rmi}$. Equation~\eqref{pos} also requires that for arbitrary $a_\rmi (x) $
\be \label{posi1}
f^{\rmi \rmj} (\chi_r)  a_\rmi (x) a_\rmj (x) \geq 0 \
\ee
which in turn implies that the dissipative coefficients in~\eqref{ii2} are non-negative and the propagators $\vev{\chi_{r \rmi} \chi_{a\rmj}}$ are retarded.
The total derivative term in (\ref{344}) in this case is
\be \begin{split}\label{vmu}
V^0_0 = &- \beta_0 F  + \cdots, \\
{1 \ov \beta_0} V^i_0 =& {\p F \ov \p \p_i \chi_{r \rmi}} \p_0 \chi_{r \rmi} + {\p F \ov \p \p_i^2 \chi_{r \rmi}} \p_i \p_0 \chi_{r \rmi}\\
 &-  \p_i {\p F \ov \p \p_i^2 \chi_{r \rmi}} \p_0 \chi_{r \rmi}  + \cdots \
\end{split}\ee
with $V_1^\mu=0$ to the order of derivative considered.

We can now readily write the entropy current to the order exhibited by applying equation~\eqref{endn}, which gives
\be
s^\mu = V^\mu_0 \ .
\ee
One can readily check that after using equations of motion
 \be
  \p_\mu s^\mu =  \beta_0^2 f_{\rmi \rmj}  \p_0 \chi_{r \rmj} \p_0 \chi_{r \rmi} \geq 0 \  .
 \ee
At zeroth order in time derivatives we have
\be
s^0 = - \beta_0 F, \qquad s^i = 0 \
\ee
which has the standard form with $F$ interpreted as the (static) free energy density of the scalar system.

Note that from our discussion of~\eqref{nemp}--\eqref{nemp1}, $\sL_1$ as given in~\eqref{ii0}--\eqref{ii1} is non-dissipative and can be factorized as
\be\begin{split}
I_1 = \int d^d x \, \sL_1 =& - \int d^d x\, F (\chi_1, \p_i \chi_1, \cdots)\\
 &+ \int d^d x \,  F (\chi_2, \p_i \chi_2, \cdots) \ .
\end{split}\ee
Of course at this order the non-dissipative statement is almost triviality as $F$ does not contain any time derivatives.

\section{EFTs  II:  a theory of diffusion} \label{sec:eft2}

In the presence of conserved quantities, the formulation of an EFT has new elements. The first issue to address
is how to identify collective degrees of freedom with conserved quantities from first principle.
There are also additional symmetries one has to impose and new elements associated with imposing the dynamical KMS symmetry.  In this section we shall illustrate the basic idea in the simplest case: a single conserved current $J^\mu$ associated with some global $U(1)$ symmetry at a fixed inverse temperature $\b_0$, i.e. we ignore disturbances in energy and momentum.



We would like to identify the collective variable(s)  associated with conservation of $J^\mu$
in a universal manner, without relying any phenomenological assumptions or details of specific systems.
For this purpose let us consider the generating functional of the conserved current along the CTP
\be\label{1gent1}
e^{ W[A_{1\mu} , A_{2 \mu}]} =
\Tr \le(\rho_0 \sP e^{i \int d^d x \, A_{1\mu} J_1^\mu -  i \int d^d x \, A_{2\mu} J_2^\mu} \ri),
\ee
where  $\rho_0$ denotes a thermal state of inverse temperature $\b_0$, and $A_{1\mu}$ and $A_{2\mu}$ are external vector fields which act as sources for the two copies of the current $J_1^\mu$ and $J_2^\mu$, respectively. The advantage of considering $W$ is that the conservation of $J^\mu$ can now be converted into a ``symmetry'' statement of $W[A_1, A_2]$. Namely,  given that $J_{1,2}^\mu$ are conserved, we have
\be \label{1shifts}
W [A_{1 \mu} , A_{2 \mu}] =W [A_{1 \mu} + \p_\mu \lam_1 , A_{2 \mu} + \p_\mu \lam_2] \quad
\ee
for arbitrary functions $\lam_1, \lam_2$,\footnote{We take $\lam_{1,2}$ to vanish at spacetime infinities.} i.e. $W$ is invariant under independent  ``gauge'' transformations of $A_{1\mu}$ and $A_{2\mu}$.

In order to obtain an effective action of collective variables associated with $J_{1,2}^\mu$, we would like to ``integrate them in,'' i.e. write $W [A_{1\mu} , A_{2 \mu}]$
as
\be \label{1newc1}
e^{W [A_{1\mu} , A_{2 \mu}]} = \int D \vp_1 D \vp_2 \, e^{i I_{\rm EFT} [\vp_1, \vp_2; A_{1\mu},  A_{2 \mu}]} ,\
\ee
where $\vp_{1,2}$ are currently place holders, 
whose nature we will elucidate in a moment. 
$I_{\rm EFT}$ should be such that: (i) Equation~\eqref{1shifts} is satisfied regardless of details of specific systems; (ii) Equations of motion of $\vp_{1,2}$ should be equivalent to conservations of $J_{1,2}^\mu$. These two conditions essentially fix the nature of $\vp_{1,2}$ completely: they should be
scalar fields and they should always appear with external fields through the combinations
\be\label{1Bvar}
B_{1 \mu}  \equiv A_{1 \mu} + \p_\mu \vp_1, \qquad B_{2 \mu} \equiv A_{2 \mu} + \p_\mu \vp_2 \ .
\ee
In other words, $\vp_{1,2}$ are the Stueckelberg fields associated with the ``gauge'' symmetries~\eqref{1shifts}. As a
result~\eqref{1newc1} can be written as
\be \label{1newc}
e^{W [A_{1\mu} , A_{2 \mu}]} = \int D \vp_1 D \vp_2 \, e^{i I_{\rm EFT} [B_{1\mu},  B_{2 \mu}]} \ .
\ee
By construction, $B_{1,2\mu}$ and so the action, are \emph{invariant} under the following transformations,
\be
A_{1,2\mu}\to A_{1,2\mu} -\p_\mu\lambda_{1,2},\qquad \varphi_{1,2}\to\varphi_{1,2}+\lambda_{1,2}\ .
\ee
The integrations over $\vp_{1,2}$ then remove the longitudinal part of $A_{1,2 \mu}$, and thus $W$ obtained from~\eqref{1newc} automatically satisfies~\eqref{1shifts}.

Now define the ``off-shell hydrodynamic'' currents as
\be \label{offc}
\frac{\delta I_{\rm EFT}}{\delta A_{1\mu} (x)}\equiv \hat J_1^\mu(x),\qquad
\frac{\delta I_{\rm EFT} }{\delta A_{2\mu}(x)}\equiv - \hat J_2^\mu(x)\ .
\ee
It then immediately follows from~\eqref{1newc} that
equations of motion for $\varphi_{1,2}$ are equivalent to the conservation equations for $\hat J^\mu_{1,2}$, i.e.
\be
{\de I_{\rm EFT} \ov \de \vp_{1,2} (x)} = - \p_\mu
 \hat J_{1,2}^\mu (x)=0\ .
\ee
Note that correlation functions of currents $J_{1,2}^\mu$ for the full theory are given by those of $\hat J_{1,2}^\mu$ in the effective field theory~\eqref{1newc}. For example,
\be\begin{split} \label{gbh}
&\langle \sP J_1^\mu(x)J_2^\nu (y)\rangle= - {\de W \ov \de A_{1 \mu} (x) \de A_{2 \mu} (y)} \biggr|_{A_1 = A_2 =0} \\
&=  \int D\varphi_1 D\varphi_2 \, e^{iI_{\rm EFT} [\p_\mu \vp_{1},\p_\mu \vp_{2}]} \, \hat J_1^\mu(x) \hat J_2^\nu(y)\ .
\end{split}\ee

So far the discussion is very general, and can in principle apply to any systems,  zero temperature or finite temperature, normal fluids or superfluids. In general, $I_{\rm EFT}$ is nonlocal. It can be considered as a mathematical device to automatically capture whatever constraints coming from current conservation. Now as discussed in the Introduction, for a generic system at a finite temperature, the only relevant slow variables are associated with conserved quantities. In this case (when we ignore other conserved quantities), the only source of non-locality of $W$ at large distance and time scales must come from integrating over $\vp_{1,2}$, thus we expect $I_{\rm EFT}$ has a well defined {\it local} derivative expansion, with the effective expansion parameter $\ell \p_\mu \sim {\ell \ov \lam} \ll 1$, where $\ell$ is the relaxation scale discussed in Sec.~\ref{sec:1b} and $\lam$ is the typical wave length of macroscopic processes we are interested in. One could also choose not to perform derivative expansion, then $I_{\rm EFT}$ is non-local, with non-locality controlled by relaxation scale $\ell$.\footnote{In contrast for a zero temperature system, there exist in general other gapless modes. To obtain $I_{\rm EFT}$, they are integrated out. In that case $I_{\rm EFT}$ is non-local to arbitrary long distance and time scales, i.e. there is no well-defined derivative expansion.}

Now let us restrict to a finite temperature system (with no other conserved quantities), and assume that the system is
in a liquid phase. There is still a distinction of a normal phase, and a superfluid phase where the $U(1)$ symmetry is spontaneously broken. It is interesting that if one directly writes down the most general local derivative expansion of $I_{\rm EFT}[B_1, B_2]$, the theory describes a superfluid phase. To describe a normal phase  one needs to impose a further symmetry, as follows.

Physically we can view the system as a continuum of fluid elements, and interpret $B_{s \mu}$ ($s=1,2$) as the ``local'' external sources for the fluid elements, which include not only external fields $A_{s \mu}$, but also
contributions from dynamical variables $\vp_{s}$. For example,
we can define the {\it local} chemical potentials as
\be\label{610}
\mu_{s} (x) = B_{s0} (x) ,\qquad s=1,2 \ . 
\ee
Now
 to describe a system for which the $U(1)$ symmetry is not spontaneously broken, we require $I_{\rm EFT}$ to be invariant under a {\it time-independent}, diagonal gauge transformations of $B_{1,2\mu}$ (to which we will refer as the diagonal shift symmetry)
 \be \begin{split}\label{daug}
 B_{1i}& \to  B_{1i}' = B_{1i} - \p_i \lam (x^i), \\
 B_{2i}& \to  B_{2i}' = B_{2i} - \p_i \lam (x^i), \
\end{split} \ee
or equivalently
\be\begin{gathered}\label{cshift}
\vp_r \to \vp_r - \lam (x^i) , \quad \vp_a \to \vp_a  , \\
 \vp_r = \ha (\vp_1 + \vp_2), \quad \vp_a = \vp_1 - \vp_2 \ .
\end{gathered}\ee
The origin of~\eqref{daug} can be understood as follows. Given the $U(1)$ symmetry,
each fluid element should have the freedom of making a phase rotation. As we are considering a global symmetry,
the phase cannot depend on time $t$, but since fluid elements are independent of one another,  they should have the freedom of making independent phase rotations, i.e. we should allow phase rotations of the form
$e^{i \lam (x^i)}$, with $\lam (x^i)$ an arbitrary function of $x^i$ only. As $B_{sa}$ are the ``gauge fields'' coupled to charged fluid elements, thus the system should be
invariant under~\eqref{daug}. When the $U(1)$ symmetry is spontaneously broken, i.e. the system in a superfluid phase, the phase freedom for all fluid elements are locked together, and~\eqref{daug} should be dropped.

Let us now consider the dynamical KMS symmetry. From~\eqref{1newfdt1}--\eqref{tiV0} the generating functional should satisfy
\be
W [A_1, A_2] = W [\tilde A_1 , \tilde A_2]
\ee
with
\be\begin{split} \label{tiV9}
\tilde A_{1\mu} (x)& = \Th A_{1 \mu} (t - i \th, \vx) , \\
\tilde A_{2 \mu} (x) &=  \Th A_{2 \mu} (t + i (\beta_0 - \th), \vx ) \ .
\end{split}\ee
To achieve this, we require the action $I_{\rm EFT}$ to satisfy
\be \label{nnk}
I_{\rm EFT} [B_1, B_2] = I_{\rm EFT} [\tilde B_1 , \tilde B_2]
\ee
with
\be\begin{split} \label{tiV3}
\tilde B_{1\mu} (x) &= \Th B_{1 \mu} (t - i \th, \vx) , \\
\tilde B_{2 \mu} (x) &=  \Th B_{2 \mu} (t + i (\beta_0 - \th), \vx ) \
\end{split}\ee
which in turn requires
\be\begin{split} \label{tiV4}
\tilde \vp_{1} (x) &= \Th \vp_{1 } (t - i \th, \vx) , \\
\tilde \vp_{2 } (x) &=  \Th \vp_{2 } (t + i (\beta_0 - \th), \vx ) \  .
\end{split}\ee
In the classical limit~\eqref{tiV3}--\eqref{tiV4} become
\be\begin{split}
\tilde B_{r \mu} (x) &= \Th B_{r \mu}  (x) , \\
\tilde B_{a \mu} (x) &= \Th B_{a \mu}  (x) + i \b_0  \Th \p_t  B_{r \mu} (x),  \\
\tilde \varphi_r(x)&=\Theta\varphi_r(x),\\
\tilde \varphi_a(x) &=
\Theta\varphi_a(x) +i \beta_0 \Theta  \p_t \varphi_r(x) \ .
\end{split}\ee
See Appendix~\ref{app:c} for how $A_\mu$ and $\vp$ transform under various choices of $\Th$.

To summarize, in order to write down the effective theory for slow variables corresponding to a conserved current in a normal phase, we need to impose on $I_{\rm EFT} [B_1, B_2]$ the following conditions: (i) Diagonal shift symmetry~\eqref{daug};
(ii)~\eqref{fer1}--\eqref{1keyp2}; (iii)~\eqref{nnk}; (iv) Rotational and translation symmetries.

As an illustration here we quote the final action at quadratic order in $B_{r,a}$, and to linear order in derivatives
\be \label{1yue}
\sL_{\rm EFT} = 
i {\sig \ov \b_0}  B_{ai}^2 + \chi  B_{a0} B_{r0}
 - \sig B_{ai} \p_0 B_{ri}  ,
\ee
with $\sig \geq 0$ and $\chi$ constants.  The off-shell currents~\eqref{offc} have the form
\bega \label{1ccu}
\hat J^0_r =  \chi \mu   , \qquad
\hat J^i_r =  \sig (E_i - \p_i \mu) + i {\sig \ov \b_0} B_{ai}, \\
J_a^0 = \chi B_{a0} , \qquad J_a^i = \sig \p_0 B_{ai}
\end{gather}
where we have introduced local chemical potential $\mu = B_{r0} = A_{r0} + \p_0 \vp_r$ and background electric field
$E_i = \p_i A_{r0} - \p_0 A_{ri}$. Clearly $\sig$ corresponds to conductivity and $\chi$ to charge susceptibility.

The equations of motion are simply the conservation equations
$\p_\mu \hat J^\mu_{r,a} = 0$.
In the absence of unphysical sources, $A_{a\mu} =0$, we have $\vp_a =0$ from $\p_\mu \hat J_a^\mu =0$. We
then find that $\hat J^\mu_a =0$ and $B_{ai} =0$. The conservation equation $\p_\mu \hat J_r^\mu =0$ can then be written as
\be
\p_0 n - D \p_i^2 n = - \sig \p_i E_i, \qquad n =\hat J^0_r
\ee
where  the diffusion constant $D$ is given by
\be
 D = {\sig \ov \chi}  \ .
\ee

Note that in~\eqref{1ccu} at leading order in the $a$-field expansion $\hat J^\mu_r$ are expressed in terms of $\mu$ and
$E_i$, i.e. $B_{ri}$ does not appear. This is not an accident and in fact persists to all derivative orders and nonlinear level.
The diagonal shift symmetry~\eqref{daug} means that $B_{ri}$ can only appear either with a time derivative $\p_0 B_{ri} =-  E_i + \p_i \mu$ or through $F_{rij} = \p_i B_{rj} - \p_j B_{ri} = \p_i A_{rj} - \p_j A_{ri}$.

The above discussion can be generalized to all derivative orders and also nonlinear in $B$'s, see Sec.~IV of~\cite{CGL}.
See also~\cite{ping,yarom} for a superspace formulation.

Applying the discussion of Sec.~\ref{sec:entropy} to~\eqref{1yue}, one finds (with external fields turned off)
\be\begin{gathered} V_0^0=\frac 12\beta_0\chi \mu^2,\qquad V_0^i=0,\\
\hat V_1^0= \beta_0\chi \mu^2,\qquad \hat V^i_1=
- \sigma\beta_0\mu \p_i\mu
\ .\end{gathered}\ee
and from (\ref{endn}) we obtain the entropy density and flux
\be
\label{entdi}
s^0 = -\frac 12 \beta_0 \chi\mu^2,\qquad s^i= \sigma\beta_0 \p_i\mu\ ,    
\ee
One can verify using equations of motion that 
\be \p_\mu s^\mu = \beta_0\sigma (\p_i\mu)^2 
\geq 0\ .\ee

The non-dissipative regime can be obtained by setting $\sig =0$. We find that the resulting action can be factorized as
\be\begin{gathered}
\sL = \chi B_{r0} B_{a0} = \sL_0 (B_1) - \sL_0 (B_2), \\ \sL_0 (B) = \chi (A_0 - \p_0 \vp)^2  \ .
\end{gathered}\ee

\section{EFT III:  action for hydrodynamics} \label{sec:hydroeft}

Let us now consider the effective field theory for collective variables corresponding to conservations of energy and momentum. The resulting theory gives fluctuating hydrodynamics. Traditional formulation of hydrodynamics is based on phenomenological equations of motion which we briefly review in Appendix~\ref{app:hydro}. Here we shall formulate it in an action form, from first principles based on symmetry.

Compared with examples of earlier sections, since now local energy density can vary, there is an emergent local temperature, which leads to new elements in the formulation of dynamical KMS symmetry. For illustration purpose, we will consider a relativistic system whose only conserved quantities are the stress tensor, and only present the basic ideas and formalism.
For a more complete treatment  and generalization to a charged fluid readers should consult the original papers~\cite{CGL,CGL1}.

The search for an action principle for fluids has a long history, dating back at least to~\cite{Herglotz}
and subsequent work including~\cite{Taub,Salmon}  (see~\cite{soper,Jackiw:2004nm,Andersson:2006nr} for reviews), all of which were for ideal fluids. Recent interests in this problem started with~\cite{Dubovsky:2005xd} where the ideal fluid formulation of~\cite{Herglotz} was rediscovered and extended in various ways (see also~\cite{Dubovsky:2011sj,Dubovsky:2011sk,Nicolis:2013lma,Endlich:2010hf,Nicolis:2011ey,Nicolis:2011cs,Delacretaz:2014jka,Geracie:2014iva,Haehl:2013hoa,Haehl:2013kra}). These works made it clear that the Lagrange type variables are natural for an action principle formulation of hydrodynamics (see also~\cite{Nickel:2010pr,deBoer:2015ija,Crossley:2015tka} for discussions in
holography). The first attempts to generalize the formalism of~\cite{Dubovsky:2005xd}  to a doubled version in the closed time path formalism to include dissipation were made in~\cite{Grozdanov:2013dba,Endlich:2012vt}.
More recent works are~\cite{Kovtun:2014hpa,Harder:2015nxa,Haehl:2015pja,Haehl:2015uoc} which have some overlaps with our formulation. Further developments have been pursued in \cite{Haehl:2016pec,Geracie:2017uku,Haehl:2018lcu,Haehl:2018uqv,yarom,Jensen:2018hhx,Jensen:2018hse}.


\subsection{Fluid spacetime formulation}\label{sec:hydro1}

We would like to first identify the collective variables associated with conservation of the stress tensor $T^{\mu \nu}$ of a system from first principle. The idea is very similar to that of Sec.~\ref{sec:eft2} for a conserved current.

Turning on external sources for the stress tensor corresponds to putting the system in a curved metric $g_{\mu \nu}$.
Thus the generating functional for the stress tensor on a CTP has the form
\be\begin{split} \label{genM}
&e^{W[g_{1\mu\nu},g_{2\mu\nu}]}\\
&=\text{Tr}\,\left[U(+\infty,-\infty;g_{1\mu\nu})\rho_0 U^\dag(+\infty,-\infty;g_{2\mu\nu})\right]
\end{split}\ee
where $U(t_2,t_1;g_{\mu\nu})$ denotes the quantum evolution operator of the system from time $t_1$ to time $t_2$ in the presence of spacetime metric $g_{\mu\nu}$. From the conservation of $T_{1,2}^{\mu \nu}$,  the generating functional is invariant under independent diffeomorphisms acting on $g_{1\mu\nu}$ and $g_{2\mu\nu}$,\footnote{For simplicity, we will restrict to systems without gravitational anomalies. For treatment of systems with anomalies see~\cite{Glorioso:2017lcn}.}
\be \label{wardt}
W[g_1,g_2]=W[g_1^{\xi_1},g_2^{\xi_2}]\ ,
\ee
where $g^\xi$ denote diffeomorphisms of $g$ with parameters $\xi^\mu$, e.g.\footnote{$\xi^{\mu}_{1,2}$ are assumed to vanish at spatial and time infinities.}
\be
g_{1 AB}^{\xi_1} (\sig) = g_{1 \mu \nu}  (\xi_1 (\sig)) {\p \xi^\mu_1 \ov \p \sig^A}   {\p \xi^\nu_1 \ov \p \sig^B}   \ .
\ee
Now following exactly the same logic as the discussion around equations~\eqref{1newc1}--\eqref{1newc} in Sec.~\ref{sec:eft2},
we obtain collective variables for the stress tensor  by promoting diffeomorphism parameters $\xi_1^\mu, \xi_2^\mu$ to dynamical variables. More explicitly, we introduce dynamical variables $X_{1,2}^\mu(\sigma^A)$, and write
the generating functional~\eqref{genM} as 
\be \label{2med}
e^{W[g_{1\mu\nu},g_{2\mu\nu}]}=\int DX_1 DX_2 \, e^{ iI_{\rm EFT} [h_{1AB},h_{2AB}]}\ ,
\ee
where
\be\begin{split} \label{stuch}h_{1AB}(\sigma)&=\frac{\p X_1^\mu}{\p\sigma^A} g_{1\mu\nu}(X_1)\frac{\p X_1^\nu}{\p \sigma^B},\\ h_{2AB}(\sigma)&=\frac{\p X_2^\mu}{\p\sigma^A} g_{2\mu\nu}(X_2)\frac{\p X_2^\nu}{\p \sigma^B}
\end{split}\ee
are counterparts of $B_1, B_2$ in~\eqref{1newc} for diffeomorphisms.

\begin{figure*}[!ht]
\begin{center}
\includegraphics[width=13.5cm]{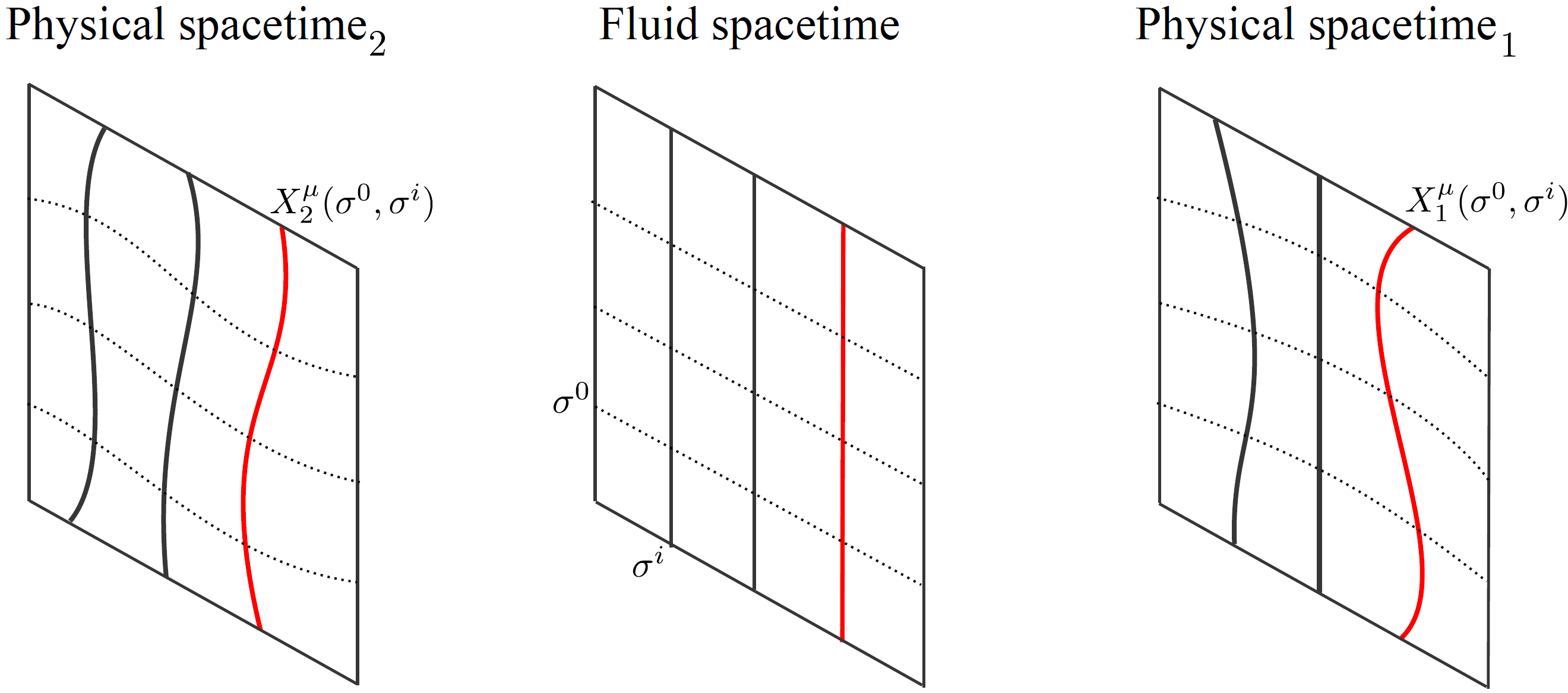} \quad
\end{center}
\caption{Relations between the fluid spacetime and two copies of physical spacetimes.  The red straight
line in the fluid spacetime with constant $\sig^i$ is mapped by $X^\mu_{1,2} (\sig^0, \sig^i)$ to
physical spacetime trajectories (also in red) of the  corresponding fluid element.
 }
 \label{fig:spaces}
\end{figure*}

In order to promote diffeomorphism parameters to dynamical variables we need to introduce a new auxiliary space-time with coordinates $\sigma^A, \; A=0,1,\dots,d-1$, to which we refer as the ``fluid space-time'' and whose interpretation will be given momentarily.  $X_1^\mu$ and $X_2^\mu$ are the coordinates of the two copies of physical space-time, where the background metrics $g_{1\mu\nu}$ and $g_{2\mu\nu}$ live. Imagining the system as a continuum of ``fluid'' elements\footnote{As will be commented below, at this stage our discussion is completely general, not necessarily restricted to fluid systems. Here we use term ``fluid'' for later convenience.}, 
it appears natural to interpret  the spatial part $\sig^i$ of $\sig^A$ as labels for fluid elements, while the time component $\sig^0$ serves as an ``internal clock'' carried by a fluid element. In this interpretation, $X_{1,2}^\mu (\sig^A)$ then corresponds to the Lagrangian description of a continuous medium. With a fixed $\sig^i$, $X^\mu_{1,2} (\sig^0, \sig^i)$  describes how a fluid element labeled by $\sig^i$ moves in (two copies of) physical spacetime as the internal clock $\sig^0$ changes.
The relation between $\sig^A$ and $X_{1,2}^\mu (\sig)$ is summarized in Fig.~\ref{fig:spaces}.
With this interpretation,
then
\be
- d \ell^2_s = g_{s \mu \nu} {\p X^\mu_s \ov \p \sig^0} {\p X^\nu_s \ov \p \sig^0} (d \sig^0)^2, \quad s=1,2
\ee
are the proper time square of the respective motions, and the corresponding velocities are given by
\be \begin{gathered}\label{umu}
u^\mu_s (\sig) = {\de X^\mu_s \ov \de \ell_s} = {1 \ov b_s} {\p X^\mu_s \ov \p \sig^0}, \\ b_s = \sqrt{- {\p X^\mu_s \ov \p \sig^0} g_{s\mu \nu} {\p X^\nu_s \ov \p \sig^0}}, \quad g_{s \mu \nu} u^\mu_s u^\nu_s = -1 \ .
\end{gathered}\ee

By construction, $h_{1,2AB}$ are pull-backs of the space-time metrics to the fluid spacetime, and are
invariant under
\be \label{bvar2}g_{s\mu\nu}\to g_{s\alpha\beta}\frac{\p X^\alpha_{s}}{\p X_{s}^{'\mu}} \frac{\p X^\beta_{s}}{\p X_{s}^{'\nu}},\qquad X_{s}^\mu\to X_{s}^{'\mu}=f^\mu_{s}(X_{s}^\mu)\ ,
\ee
where $s=1,2$, and which immediately implies that $W$ obtained from~\eqref{2med} satisfies~\eqref{wardt}. Introducing  the ``off-shell hydrodynamical'' stress tensors as
\be\begin{split} \label{deften}
\frac 12 \sqrt{-g_1}\hat T_1^{\mu\nu}(x)&\equiv \frac{\delta I_{\rm EFT}}{\delta g_{1\mu\nu}(x)},\\
\frac 12 \sqrt{-g_2}\hat T_2^{\mu\nu}(x)&\equiv -\frac{\delta I_{\rm EFT}}{\delta g_{2\mu\nu}(x)}\ ,
\end{split}\ee
one can readily show from the structure of $h_{1,2}$ that  equations of motion of $X_{1,2}^\mu$
are equivalent to
\be
 \nabla_{s\mu}\hat T^{\mu\nu}_{s}=0, \qquad s =1,2
 \ee
where $\nab_{s\mu}$ are the covariant derivatives associated with $g_{1,2}$ respectively.
Correlation functions of the stress tensor in the full theory are obtained from those of $\hat T_{1,2}^{\mu \nu}$
in the effective theory~\eqref{2med} in complete parallel with~\eqref{gbh}.

The comments we made after~\eqref{gbh} for the $U(1)$ case should be repeated here.
The above discussion is so far very general, applicable to any systems,  zero temperature or finite temperature, solids or liquids.
$X^\mu_{1,2}$ 
 can be considered as a mathematical device
whose dynamics automatically captures whatever constraints coming from the conservation of stress tensor.

Now let us specify to a generic finite temperature system in a fluid phase (i.e. with unbroken translational and rotational symmetries), for which
 $X^\mu_{1,2}$ are then the only relevant slow variables, and $I_{\rm EFT}
$ should have a well defined {\it local} derivative expansion, with the effective expansion parameter $\ell \p_\mu \sim {\ell \ov \lam} \ll 1$. In particular,  $I_{\rm EFT} [h_1, h_2]$ should recover the standard formulation of hydrodynamics as its equations of motion.
%
For comparison with the standard formulation,
it is convenient to introduce an additional scalar field $\tau(\sigma^A)$,
which gives the local {\it proper} temperature associated with each fluid element
\be \label{locTh}
 T(\sigma)=\frac 1{\beta(\sigma)}=T_0 e^{-\tau(\sigma)}\ ,
 \ee
where $T_0=\beta_0^{-1}$ is some reference temperature scale (say the temperature at infinity).
With the introduction of $\tau$, the path integrals~\eqref{2med} become
\be \label{3med}
e^{W[g_{1\mu\nu},g_{2\mu\nu}]}=\int DX_1 DX_2 D \tau \, e^{ iI_{\rm EFT} [h_{1AB},h_{2AB}, \tau]}\ .
\ee

Now in order to describe the system in a fluid phase,  in contrast to e.g. solids or liquid crystals, we have to impose additional symmetries. In a fluid phase, a fluid element can move freely. To reflect this we require
 that $I$ should be invariant under:
 \ben
 \item  time-independent reparameterizations of spatial manifolds of $\sig^A$, i.e.
 \be \label{sdiff}
 \sig^i \to  \sig'^i (\sig^i), \qquad \sig^0 \to \sig^0  \ ;
 \ee

 \item time-diffeomorphisms of $\sig^0$, i.e.
 \be \label{tdiff}
 \sig^0 \to \sig'^0= f (\sig^0, \sig^i), \qquad \sig^i \to \sig^i   \ .
 \ee

\een
Equation~\eqref{sdiff} corresponds to a (time-independent) relabeling of fluid elements, while~\eqref{tdiff}
can be interpreted as reparameterizations of the internal time associated with each fluid element. Note that in~\eqref{tdiff} we allow time reparameterizations to have arbitrary dependence on $\sig^i$, which physically can be interpreted as
each fluid element having its own choice of time. 
 These symmetries ``define'' what we mean by a fluid. Note that there is no covariance between $\sig^0$ and $\sig^i$ as
the fluid itself ``defines'' a frame. Two copies of physical space-time, together with an auxiliary, fluid space-time (which was called worldvolume), and similar diffeomorphisms were also discussed in~\cite{Haehl:2015pja}.


Given that the theory should be invariant under separate spatial and time diffeomorphisms (\ref{sdiff})--(\ref{tdiff}), it is convenient to decompose $h_{1,2}$ into objects which have good transformation properties under them
\be\label{613}
h_{sAB} d\sigma^A d\sigma^B=-b_s^2(d\sigma^0-v_{si}d\sigma^i)^2+a_{sij}d\sigma^i d\sigma^j \ ,
\ee
with $s=1,2$. Again symmetric combinations of various components represent physical variables, while the antisymmetric combinations are interpreted as the corresponding noises. For example, introducing
\be
 E_r =  \ha \le( b_1 +b_2 \ri) ,\quad E_a
 =   \log \le(b_2^{-1} b_1 \ri)  \
\ee
we may interpret $d \hat \sig = E_r d \sig^0$ as the local proper time for each fluid element.

From~\eqref{locTh} the local temperature (associated with $\sig^0$) for each fluid element can then be written as
\be\label{loT}
T_{\rm local} (\sig) = {T (\sig)  E_r} = T_0 E_r e^{-\tau} \ .
\ee
Note that one could use the time diffeomorphism~\eqref{tdiff} to fix
\be \label{yuep}
E_r = e^\tau
\ee
for which we have
\be \label{ye1}
T_{\rm local} (\sig) = T_0  = {\rm const} \ .
\ee
Physically this means that by properly choosing the fluid time $\sig^0$ can make
the local temperature associated with each fluid element to be a constant. In the gauge~\eqref{yuep}, eq. \eqref{tdiff}
still has some residual freedom left, reducing to
\be \label{tdiff2}
\sig^0 \to \sig^0 + f (\sig^i) \
\ee
for an arbitrary function $f$. Thus instead of having both~\eqref{sdiff} and~\eqref{tdiff}, with $\tau$ as an independent field, one can instead
have~\eqref{sdiff} and~\eqref{tdiff2}, while interpreting
\be \label{ine0}
T(\sig) = {T_0 \ov E_r}
\ee
as an emergent local {\it proper} temperature which is expressed in terms of $X^\mu_{1,2}$ through $E_r$. It can be shown that the Fadeev-Popov determinant in fixing the gauge~\eqref{yuep} in the path integral~\eqref{3med} is unity
and thus the two formulations are equivalent at full path integral level.


Finally we need to specify the dynamical (local) KMS symmetry. Let us first suppose $\rho_0$ in~\eqref{genM} is given by a thermal density matrix with inverse temperature $\b_0 = {1 \ov T_0}$. Then the generating functional~\eqref{genM} should satisfy
\be \label{osfdt}
W[g_{1\mu\nu},g_{2\mu\nu}]=W[\tilde g_{1\mu\nu},\tilde g_{2\mu\nu}]\ ,
\ee
with
\be\begin{split}\label{hdkms1} \tilde g_{1\mu\nu}(x)&=\Th g_{1\mu\nu}(t-i\theta_0,\vec x),\\
\tilde g_{2\mu\nu}(x)&= \Th g_{2\mu\nu}(t+i(\beta_0-\theta_0),\vec x)\ .
\end{split}\ee
Compared with~\eqref{tiV1} and~\eqref{tiV3}, the additional complication here is that the dynamical KMS condition must be specified with respect to the local temperature, which is now spacetime dependent.
In the gauge~\eqref{ye1}, the expression simplifies and the dynamical KMS transformation particularly is easy to state.\footnote{See Sec. IV of~\cite{CGL1} for a general discussion.} We require $I_{\rm EFT}$ to be invariant under
\be \label{bnh}
I_{\rm EFT} [\tilde h_1, \tilde h_2] = I_{\rm EFT} [ h_1,  h_2]
\ee
with
\be \begin{split}\label{hdkms2}
\tilde h_{1AB}(\sig)&=\Th h_{1AB}(\sigma^0 - i \th, \sig^i),\\
\tilde h_{2AB}(\sigma)&=\Th h_{2AB}(\sigma^0+i (\beta_0 - \th), \sig^i)\ .
\end{split}\ee
 Equations~\eqref{hdkms1} and~\eqref{hdkms2} in turn imply that $X_{1,2}^\mu$ should transform as
\be\begin{split}
\label{hdkms}
\tilde X_1^\mu(\sigma) &=  \Theta X_1^\mu(\sig^0 - i \th, \sig^i) + i \th \delta^\mu_0 ,\\
\tilde X_2^\mu(\sigma)&=  \Theta X_2^\mu(\sigma^0+i(\beta_0 - \th),\sigma^i)-i(\beta_0- \th)\delta^\mu_0   \ .\end{split}
\ee
See Appendix~\ref{app:c} how various quantities transform under different choices of $\Th$.
It can be readily checked that in~\eqref{2med}, equation~\eqref{bnh} leads to~\eqref{osfdt}.\footnote{ Note a general fact: suppose an action has a symmetry
$I [\chi; \phi] = I [\tilde \chi; \tilde \phi]$
where tilded variables are related to the original variables by some transformation,
then $W  [\phi] = W  [\tilde \phi]$
where $e^{W[\phi]}=\int D\chi e^{iI[\chi;\phi]}$.}
Given the residual time diffeomorphism~\eqref{tdiff2}, in~\eqref{hdkms2}--\eqref{hdkms}, the parameter $\th$ can in fact be generalized to an arbitrary function $\th (\sig^i)$ which approaches $\theta_0$ at infinity.
 Equations (\ref{hdkms2}) and (\ref{hdkms}) are manifestly $Z_2$.

Combining all these elements and the unitarity constraints~\eqref{fer1}--\eqref{1keyp2}, we then have the complete
formulation of fluctuating hydrodynamics which is valid at quantum level. It can be checked explicitly that, at the level
of equations of motion, this formulation completely reproduces the standard formulation of hydrodynamics, with all the phenomenological
constraints such as the local first law, the local second law and Onsager relations automatically incorporated.
Furthermore, it generalizes constraints from Onsager relations to nonlinear level and provides a derivation of the relation between thermal partition function and entropy constraints observed in~\cite{Banerjee:2012iz,Jensen:2012jh}.
For details see Sec. V of~\cite{CGL}.

We will present the explicit action to first derivative order in Sec.~\ref{sec:acf} after discussing various other aspects.





\subsection{Physical spacetime formulation}  \label{sec:review}

Alternatively, we can use the inverse functions $\sig_{1,2}^A (x)$ of $X_{1,2}^\mu (\sig^A)$
as dynamical variables and rewrite the fluid spacetime action in the physical spacetime.
Now there is only a single copy of physical spacetime $x^\mu$ as the arguments
of $\sig^A_{1,2}$ are dummy variables. 
 The background metrics are $g_{1 \mu \nu} (x), g_{2 \mu \nu} (x)$.

With general background sources, the action appears to be complicated and not very transparent written in physical spacetime. A reason for the complications is the following. In the fluid spacetime we can use the symmetric combination of the induced metric
\be
h_{r AB} = \ha (h_{1 AB} + h_{2 AB})
\ee
as the ``physical metric'' and use it to raise and lower indices. This makes sense as $h_{1,2}$ have the same transformation properties under~\eqref{sdiff}--\eqref{tdiff}. But there does not exist a canonical definition of the ``physical'' spacetime metric. The obvious candidate   $g_{\mu \nu} = \ha (g_{1 \mu \nu} + g_{2 \mu \nu})$ does not make sense
as $g_1$ and $g_2$  transform under independent diffeomorphisms~\eqref{bvar2}. Thus it is not a good thing to add them.
This problem does not exist in the classical limit as we will see in next subsection.

\subsection{Classical limit}  \label{sec:classical}

So far the discussion applies to any quantum system and includes quantum fluctuations. In this section we consider the classical limit $\hbar\to 0$, which simplifies the structure of the hydrodynamical action as well as the dynamical KMS transformation.
Now there is a canonical physical spacetime metric (even when $g_1 \neq g_2$), and the fluid spacetime and physical spacetime quantities--including actions--are now simply related by pullbacks.



Following the discussion of Sec.~\ref{sec:hbar}, reinstating $\hbar$ we can write various background and dynamical fields as
\begin{align}
g_{1\mu\nu}=&g_{\mu\nu}+\frac  \hbar2 g_{a\mu\nu} ,&\qquad g_{2\mu\nu} =&g_{\mu\nu}-\frac \hbar2  g_{a\mu\nu},\notag \\
X_1^\mu =&X^\mu+\frac \hbar2   X_a^\mu ,&\qquad  X_2^\mu =&
X^\mu-\frac \hbar 2  X_a^\mu , \label{re1a}
\end{align}
where now $g_{\mu \nu}$ and $X^\mu$ are interpreted as the physical spacetime metric and coordinates~(note there is only one copy of them).
In~\eqref{bvar2}, the transformation parameters $f_{1,2}^\mu$ can be written as
\begin{align}
f_1^\mu=f^\mu+\frac 12 \hbar  f_a^\mu ,\qquad  f_2^\mu=f^\mu-\frac 12 \hbar f_a^\mu \ .
\end{align}

In the  $\hbar \to 0$ limit, the two diffeomorphisms~\eqref{bvar2} then become: (i)  physical space diffeomorphisms
\be\label{ddeo}
X^\mu \to X'^\mu (X) = f^\mu (X), 
\ee
under which $X^\mu_a$ transform as a vector and $g_{\mu \nu}, g_{a \mu \nu}$ as symmetric tensors, and (ii) noise diffeomorphisms under which various quantities transform as
\be \label{upo}
X_a'^\mu (\sig) = X_a^\mu (\sig) + f_a^\mu (X (\sig)), \quad g_{a \mu \nu}'   = g_{a \mu \nu} - \sL_{f_a} g_{\mu \nu}   \ ,
\ee
where $\sL_w$ denotes Lie derivative along a vector $w^\mu$.
We emphasize that~\eqref{upo} are finite transformations. They are exact as $\hbar\to 0$, and do not have derivative corrections.
Note that $g_{\mu \nu}$ is naturally interpreted as the physical spacetime metric.

In this limit
\be\begin{split}
h_{1AB}  
&= h_{AB} (\sig) +  {\hbar \ov 2} h_{aAB}  , \\
h_{2AB}  
&= h_{AB} (\sig) -  {\hbar \ov 2} h_{aAB},
\label{b1}\end{split}
\ee
where
\begin{eqnarray}\label{pp0}
h_{AB} (\sig) &\equiv&  \p_A X^\mu\p_B X^\nu g_{\mu \nu} (X) , \\
h_{aAB}(\sig) &=& \p_A X^\mu\p_B X^\nu G_{a \mu \nu} (X), \label{pp1}\\
G_{a\mu\nu} (X) 
& \equiv& g_{a\mu \nu} + \sL_{X_a} g_{\mu \nu}
\ .
\label{pp2}
\end{eqnarray}
It can be readily checked that $G_{a \mu \nu}$ is invariant under~\eqref{upo} and transforms as a symmetric tensor under~\eqref{ddeo}.  We now have
\be
{1 \ov \hbar} I_{\rm EFT} [h_1, h_2, \tau]  = I_{\rm EFT} [h_{AB}, h_{aAB},  \tau] + O(\hbar) \ .
\ee
One also has a natural definition of ``physical'' velocity field (rather than two copies of them as in~\eqref{umu}) as
\be \label{0vel}
u^\mu (\sig)= {1 \ov b} {\p X^\mu \ov \p \sig^0} , \qquad b = \sqrt{- h_{00}} \ .
\ee

Going to physical spacetime formulation, we  treat $\sig^A (x^\mu)$, the inverse of $X^\mu (\sig^A)$, and
$X^\mu_a (x) = X^\mu_a (\sig^A (x))$ (now understood as a vector in the physical spacetime) as
dynamical variables. Similarly, the velocity field~\eqref{0vel} should now understood as defined in physical spacetime through $\sig^A (x)$, more explicitly,
\be\begin{gathered} \label{1vel}
u^\mu (x) ={1 \ov b}  (K^{-1})^\mu_0 , \qquad K_\mu^A = \p_\mu \sig^A , \\ b^2 = - g_{\mu \nu}  (K^{-1})^\mu_0  (K^{-1})^\nu_0 \ .
\end{gathered}
\ee
Invariance under~\eqref{sdiff}--\eqref{tdiff} implies that
the only invariant which can be constructed from $K_\mu^A$ is the velocity field
$u^\mu$ and the projector to directions transverse to $u^\mu$
\be
\De^{\mu \nu} = g^{\mu \nu} + u^\mu u^\nu \ .
\ee
Thus the invariance under~\eqref{upo} and~\eqref{sdiff}--\eqref{tdiff} implies that the only variables which can appear in the action
of physical spacetime are:
\be \label{dvar}
G_{a \mu \nu}, \quad u^\mu, \quad \De^{\mu \nu}, \quad \b (x) = \b (\sig (x))
\ee
and their covariant derivatives. The action should also be invariant under physical spacetime diffeomorphisms~\eqref{ddeo}.
The fluid spacetime action is obtained by that of the physical spacetime by pulling back all quantities to the fluid spacetime.

To discuss the classical limit of the dynamical KMS transformation it is convenient to introduce the combination
\be
\b^\mu (x) = \b (x) u^\mu , \qquad   \b (x) = \b_0 {e^\tau }
\ee
and its pull back in fluid space
\be\begin{split}
\b^A (\sig) &= \b^\mu {\p \sig^A \ov \p x^\mu}\\
 &= \b_0 {e^\tau \ov b} \le({\p \ov \p \sig^0} \ri)^A
= \b_{\rm local} \le({\p \ov \p \sig^0} \ri)^A
\end{split}\ee
where we have used~\eqref{loT} and that in the classical limit  $E_r = b$.
In the gauge~\eqref{ine0}, we have 
\be
\b^\mu (x) = \b_0 {\p X^\mu \ov \p \sig^0} = \b_0 b  u^\mu , \quad \b^A  (\sig) = \b_0 \le({\p \ov \p \sig^0} \ri)^A   \ .
\ee
In the gauge~\eqref{ine0}, the remaining symmetry~\eqref{tdiff2} is no longer enough to reduce $\sig^A$ to $u^\mu$ as is the case for~\eqref{tdiff}, there is one more invariant variable $b$. Thus in physical spacetime, $\b(x)$ is always treated
an independent variable.

In the  $\hbar \to 0$ limit~\eqref{re1a}, with $\th, \beta_0$ in~\eqref{hdkms1}--\eqref{hdkms} becoming $\hbar \th, \hbar \beta_0$.
those equations can be written as
\begin{eqnarray} \label{hdckms}
\tilde g_{\mu\nu}(x)&=& \Th g_{\mu\nu}(x),\\
 \tilde g_{a\mu\nu}(x)&=& \Th g_{a\mu\nu}(x)+i\beta_0\Theta\p_0 g_{\mu\nu}(x)\\
 \label{hdckms1}
 \tilde X^\mu(\sigma)&=&\Th X^\mu(\sigma),\\
  \tilde X_a^\mu(\sigma)&=& \Th X_a^\mu(\sigma)-i\Theta\beta^\mu(\sigma)+i\beta_0 \delta^\mu_0\ , \\
  \label{hdckms2}
  \tilde h_{ab}(\sigma)&=&\Th h_{ab}(\sigma),\\
   \tilde h^{(a)}_{ab}(\sigma)&=&\Th h^{(a)}_{ab}(\sigma)+i\Theta\mathcal L_\beta h_{ab}(\sigma)\ .
\end{eqnarray}
In fact it can be shown that~\eqref{hdckms1}--\eqref{hdckms2} are valid without choosing the gauge~\eqref{ine0}~\cite{CGL1}.

Using (\ref{hdckms})--(\ref{hdckms2}) together with the pull-backs (\ref{pp0})-(\ref{pp1}), we find
the dynamical KMS  transformation for physical spacetime quantities
\be\label{dynkh}\begin{gathered}
\tilde u^\mu(x)=\Th u^\mu(x),\quad \tilde \beta(x)=\Th \beta(x),\\
\tilde G_{a\mu\nu}(x)=\Th G_{a\mu\nu}(x)+i\Theta\mathcal L_\beta g_{\mu\nu}(x)\ .\end{gathered}
\ee
Notice that in~\eqref{dynkh}, the dynamical KMS transformation is expressed solely in terms of a local temperature $\b (x)$.
This suggests that it can be extended to general density matrices for which the concept of a local temperature makes sense.
In fact, we can turn the logic around to  use the invariance under~\eqref{dynkh} as a mathematical definition of a local equilibrium state.

\subsection{Explicit action and field redefinitions} \label{sec:acf}

Finally let us give the explicit form of the action. We can expand the action $I_{\rm EFT}$ in terms of the number of
$a$-variables and derivatives. For simplicity we will write the action in physical spacetime in the classical limit.

We need to write down the most general covariant action using variables in~\eqref{dvar} and impose unitarity conditions~\eqref{fer1}--\eqref{1keyp2} as well as invariance under~\eqref{dynkh}. 
Writing $I_{\rm EFT} = \int d^d x \, \sqrt{-g} \, \sL$,
we will organize the Lagrangian as~\eqref{Ll}, where now the derivatives are counted as acting on variables in~\eqref{dvar}, and work to the level of $\sL_2$.\footnote{To this level, one can check that  the full quantum action in fact coincides with that in the classical limit.}

At order $\sL_1$, the most general covariant action built from~\eqref{dvar} with zero derivative is
\be \label{idac}
\sL_1 = \frac 12 T_0^{\mu\nu}G_{a\mu\nu},
\ee
with
\be \label{idst}
T_0^{\mu\nu} =  \ep_0  (\b) u^\mu u^\nu + p_0 (\b)  \De^{\mu \nu}
\ee
where for now $\ep_0, p_0$ are arbitrary functions of $\b$. Imposing invariance under~\eqref{dynkh} requires
\be \label{thermo}
\ep_0 + p_0  = - \b {\p p_0 \ov \p \b} \
\ee
which is equivalent to the local first law of thermodynamics. $\sL_2$ can be written as
\be\label{nmp}
\mathcal L_2=\frac 12 T_1^{\mu\nu}G_{a\mu\nu}+\frac i4 W_0^{\mu\alpha,\nu\beta}G_{a\mu\nu}G_{a\alpha\beta}\ ,
\ee
where $T_1^{\mu\nu}$ and $W_0$ contain first and zero derivative respectively. More explicitly, $T_1^{\mu \nu}$ can be written as
\be\label{1gens}
T^{\mu \nu}_1 = h_\ep u^\mu u^\nu + h_p \De^{\mu \nu} +
2  u^{(\mu} q^{\nu)}  - \eta \sig^{\mu \nu} \ ,
\ee
with
\be\begin{split}
 \label{1p}
 h_\epsilon &=
f_{11} \beta^{-1}\p\beta +  f_{12}  \th , \\
 h_p &=  f_{21} \beta^{-1}\p \beta - f_{22} \th  ,\\
 q^\mu &= - \lam_{1}  \p u^\mu
  + \lam_5 \De^{\mu \nu} \beta^{-1}\p_\nu \beta , \end{split}\ee
  \be\begin{gathered}
 \p \equiv u^\mu \nab_\mu, \quad \th\equiv \nab_\mu u^\mu, \\
  \sig^{\mu \nu} \equiv \De^{\mu \lam} \De^{\nu \rho} \le(\nab_\lam u_\rho + \nab_\rho u_\lam - {2 \ov d-1} g_{\lam \rho}
\nab_\al u^\al \ri) \
\label{qcur4}
\end{gathered}\ee
and
\be\begin{split}
\label{W01}
W_0^{\mu\alpha,\nu\beta}=& s_{11}u^\mu u^\nu u^\alpha u^\beta+ s_{22}\Delta^{\mu\nu}\Delta^{\alpha\beta}\\
 &- s_{12}(u^{\mu} u^{\nu}\Delta^{\alpha\beta}+u^{\alpha} u^{\beta}\Delta^{\mu\nu})\\
&+ 2 r_{11}\left(u^\mu u^{(\alpha}\Delta^{\beta)\nu}+u^\nu u^{(\alpha}\Delta^{\beta)\mu}\right)\\
&+ 4 r
\le(\Delta^{\alpha(\mu}\Delta^{\nu)\beta} - {1 \ov d-1}\Delta^{\mu\nu}\Delta^{\alpha\beta}  \ri)\ ,
\end{split}\ee
where all coefficients are functions of $\beta$. Imposing the dynamical KMS symmetry gives three sets of conditions. The first set is equivalent to requiring the existence of equilibrium, and give
\be \lambda_1=\lambda_5\ .\ee
The second set of conditions can be shown to be equivalent to the non-linear Onsager relations (\ref{1con1}), and give
\be f_{21}=-f_{12}\ .\ee
The third set of conditions relate coefficients of $T_1^{\mu\nu}$ with coefficients of $W_0^{\mu\alpha,\nu\beta}$:
\be\begin{gathered} r=\frac\eta {2\beta},\quad r_{11}=\frac{\lambda_1}{\beta},\quad  s_{11}=\frac{f_{11}}\beta,\\
s_{12}=\frac{f_{12}} \beta,\quad s_{22}=\frac{f_{22}}\beta \ .
\end{gathered}\ee

Taking derivative with respect to $g_{a \mu \nu}$ we then find the symmetric part of the off-shell hydrodynamic stress tensor is given by
\be
\hat T^{\mu \nu} =  T^{\mu \nu}_0 +  T^{\mu \nu}_1
\ee
with $T_0^{\mu \nu}$ for ideal fluids and $ T^{\mu \nu}_1$ the leading dissipative corrections.
Applying~\eqref{endn} of Sec.~\ref{sec:entropy} to the above action we find that to first order in derivative the entropy current takes the form
\be S^\mu = p \beta^\mu- T^{\mu\nu}\beta_\nu , \qquad p = p_0 + h_p
\ee
which recovers the standard result. 

To the order of $\sL_2$, the structure of~\eqref{idac} and (\ref{nmp}) parallels that of the MSR action. This was anticipated in~\cite{Kovtun:2014hpa}, where the near-equilibrium form of $\sL_2$ was obtained from the knowledge of two-point functions, and its full non-linear expression was found using Lorentz invariance. Subsequently, \cite{Harder:2015nxa} took an important step to formulate the action based on symmetries, and obtained the near equilibrium form of $W_0$. Note however that those works did not capture $\mathcal L_3$, which is important when second derivative terms in the stress tensor are relevant.





As usual, in an EFT,  one can reduce the number of terms using field redefinitions. In the current theory,
possible redefinitions are\footnote{The corresponding redefinitions for $\sig^A$ may be non-local, but this does not matter as the action only depends on $u^\mu$.}
\be\label{re1}
u^\mu\to u^\mu+\delta u^\mu ,\; \beta\to\beta+\delta\beta,\; X_a^\mu\to X_a^\mu+\delta X_a^\mu\ ,\ee
where $\delta u^\mu,\delta\beta$ are local expressions in $u^\mu,\beta$ and $G_{a \mu \nu}$ which contain  at least one derivatives or one factors of $G_{a\mu\nu}$, while $\delta X_a^\mu$ contains at least one factors of $G_{a\mu\nu}$.

The redefinitions (\ref{re1}) leave $\mathcal L_1$ invariant, but will modify $\mathcal L_l$, for $l>1$. Redefinitions of $X_a^\mu$ allow one to set to zero terms which are proportional to the ideal fluid equations of motion $\p_\mu T_0^{\mu\nu}=0$ or its derivatives, while redefinitions of $u^\mu$ and $\beta$ can be used to further simplify the action by writing it in a specific ``frame.'' 
These field redefinitions generalize those in the traditional formulation of hydrodynamics, which are used to simplify one-point function of $\hat T^{\mu\nu}$. Here the frame choice is applied to the full action, providing simplifications also for higher-point functions of the stress tensor. See~\cite{CGL1} for more detailed discussion.

In the generalized Landau frame introduced in~\cite{CGL1}, equations~\eqref{1gens}--\eqref{W01} simplify to
\begin{eqnarray} T_1^{\mu\nu}&=&-\eta\sigma^{\mu\nu}-\zeta \theta \Delta^{\mu\nu}\notag\\
\notag W_0^{\mu\alpha,\nu\beta}&=&2\beta^{-1}\eta\le(\Delta^{\alpha(\mu}\Delta^{\nu)\beta} - {1 \ov d-1}\Delta^{\mu\nu}\Delta^{\alpha\beta}  \ri) \\ &&+\beta^{-1}\zeta\Delta^{\mu\alpha}\Delta^{\nu\beta}\ ,
\end{eqnarray}
where $\eta$ and $\ze$ are shear and bulk viscosities.
In particular, the full $\sL_2$ can be written in a compact form
\be\begin{split}
\sL_2 
=&\frac i4\beta^{-1}\zeta\Delta^{\mu\nu}\hat G_{a\mu\nu}\Delta^{\alpha\beta} G_{a\alpha\beta}+\frac i2 \beta^{-1}\eta\, \times\\
&\times\le(\Delta^{\alpha(\mu}\Delta^{\nu)\beta} - {1 \ov d-1}\Delta^{\mu\nu}\Delta^{\alpha\beta}  \ri)\hat G_{a\mu\nu} G_{a\alpha\beta},
\end{split}\ee
where $\hat G_{a \mu \nu} \equiv \Th \tilde G_{a\mu \nu}$ with $\tilde G_{a \mu \nu}$ the dynamical KMS transformation~\eqref{dynkh}.

Terms at order $\sL_3$ or higher can in principle be written down straightforwardly, but
one sees a proliferation of terms, which render the analysis quite lengthy.  See~\cite{CGL1} for a discussion of conformal fluids.

The action for fluctuating hydrodynamics has been generalized in a number of directions, including systems with quantum anomalies. It has been widely  recognized that
systems with quantum anomalies exhibit novel transport behavior in
the presence of rotation or in a magnetic field (for a recent review see~\cite{Landsteiner:2016led}).
The action principle formulation provides a systematic way to derive such anomalous effects,
makes manifest the relations between parity-odd transport and underlying discrete symmetries, and elucidates connections between anomalous transport coefficients and global anomalies~\cite{Glorioso:2017lcn}.

\subsection{Ideal fluids: factorization and equivalence with single copy action}

As a support for the conjecture~\eqref{nemp1} that generic non-dissipative action should be factorizable, in this subsection we show that  the ideal fluid action~\eqref{idac} can be factorized.
Interestingly, the factorized action corresponds to  the ``single-copy'' ideal fluid action of~\cite{Dubovsky:2005xd}~(see also~\cite{Nickel:2010pr,Dubovsky:2011sj,Haehl:2015pja,Dubovsky:2011sk,Nicolis:2013lma,Endlich:2010hf,Nicolis:2011ey,Nicolis:2011cs,Delacretaz:2014jka,Geracie:2014iva,Haehl:2013hoa,Haehl:2013kra}).

For this purpose it is convenient to work in the fluid spacetime, in which~\eqref{idac} can be written as
\be\begin{split}
I_{1} &= \ha \int d^d \sig \, \sqrt{-h} \, T_0^{AB} h_{a AB} , \\
T_0^{AB} &= \p_\mu \sig^A \p_\nu \sig^B T_0^{\mu \nu}  \ .
\end{split}\ee
We would like to show that with $T_0^{\mu \nu}$ given by~\eqref{idst}--\eqref{thermo},  the above action can be written as
\be
I_1= I_0 [h_1] - I_0 [h_2]  + O(h_a^3)
\ee
for some local action $I_0$. This is equivalent to the statement that there exists $I_0 [h_{AB}]$ such that
\be
\ha \sqrt{-h} T_0^{AB} = {\de I_0 \ov \de h_{AB}}
\ee
which in turn is equivalent to the integrability condition
\be \label{unj}
{\de (\sqrt{-h} T_0^{AB}) \ov \de h_{CD}} = {\de (\sqrt{-h} T_0^{CD}) \ov \de h_{AB}} \ .
\ee
Note that we shall use the gauge~\eqref{yuep} so that $\b (\sig) = \b_0 \sqrt{- h_{00}}$.
It can be readily checked~\eqref{unj} is indeed satisfied, and $I_0$ can be written as\footnote{Note that this analysis is in fact identical to that of Sec. V F 1 of~\cite{CGL}, which worked with a charged fluid. So this analysis can be immediately generalized to a charged fluid from the results there.}
\be
\label{onep1} I_0=\int d^d \sigma\sqrt{-h}\, p_0(\b (\sig)) \ .
\ee
Note that~\eqref{onep1} is invariant under~\eqref{sdiff} and~\eqref{tdiff2}. This action is of the form discussed in~\cite{Nickel:2010pr,Haehl:2015pja} which may be considered as a covariant generalization of that in~\cite{Dubovsky:2005xd}.
To obtain the form given in~\cite{Dubovsky:2005xd}, one can write it in the physical spacetime by inverting $X^\mu (\sig)$ and
integrating out $\sig^0 (x)$, after which one obtains an action of $\sig^i (x)$. When solving $\sig^0$ in terms of other variables, one finds an arbitrary integration function, which can be fixed to be unity and in turn
breaks the symmetry~\eqref{sdiff} down
to volume-preserving spatial diffeomorphisms
\be
\sigma^i\to f^i(\sigma^j),\qquad \det \le(\frac{\p f^i}{\p \sigma^j} \ri)=1\ .
\ee

\section{Conclusion} \label{sec:conc}

In these lectures we reviewed the basic formalism for constructing EFTs for non-equilibrium systems at a finite
temperature, and discussed a few examples as illustrations, including a theory of diffusion and fluctuating hydrodynamics. The key points are:
\begin{itemize}
\item Non-equilibrium EFTs satisfy a set of universal conditions and symmetries: \emph{(a)} the constraints from unitarity (\ref{fer1})-(\ref{1keyp2}), and \emph{(b)} the $Z_2$ dynamical KMS invariance (\ref{lkms}) which characterizes local equilibrium. These, in turn, imply Onsager relations, fluctuation-dissipation theorem, existence of equilibrium and second law of thermodynamics. While some of there conditions have been traditionally imposed at phenomenological level, here they follow from first principles.

\item Systems with conserved quantities are characterized by additional symmetries. For example, charge diffusion  of Sec. \ref{sec:eft2} requires additional diagonal shift symmetries, and hydrodynamics of Sec. \ref{sec:hydroeft} requires additional diffeomorphism invariance.

\item Physical effects from thermal and quantum fluctuations can be treated systematically  by applying standard field theory methods to non-equilibrium EFTs.

\end{itemize}

As discussed in the Introduction, we expect these EFTs to have rich applications to a large variety of physical problems. The formalism also admits generalizations in many directions. For example, it would be interesting to find EFTs for other continuous media, such as solids and liquid crystals. Furthermore,
one may be able to adapt the formalism to systems where the concept of local equilibrium breaks down, but which still admit a separation of scales.
Finally, the collective degrees of freedom associated with conserved quantities are formulated in a way which does not depend on any long wavelength expansion or local equilibrium. We thus expect that they have much wider applications, e.g. to systems at very low or zero temperatures.




\section*{Acknowledgements}
We would like to thank Michael Crossley, Ping Gao and Srivatsan Rajagopal for collaborations, and
students at  ICTP SAIFR School and TASI 2017 for questions.  This work is supported by the Office of High Energy Physics of U.S. Department of Energy under grant Contract Number  DE-SC0012567. P.G. acknowledges the support of a Leo Kadanoff Fellowship, and the hospitality of Technion, where part of this work was done.

\appendix

\section{Review of standard formulation of  hydrodynamics} \label{app:hydro}

In this section we give a brief review of the standard formulation of hydrodynamics (for a more extensive, modern review see~\cite{Kovtun:2012rj}. For simplicity, we shall consider relativistic normal fluids. The equations of motion are the conservation of the stress tensor $T^{\mu\nu}$, i.e.
\be\label{cons1} \p_\mu T^{\mu\nu}=0\ ,\ee
and if there is an additional global $U(1)$ symmetry we also have the conservation of the corresponding current
\be \label{cons2} \p_\mu J^\mu=0\ .\ee

Consider first the system in thermal equilibrium, where the density matrix $\rho_0$ has the general form
\be \rho_0=\frac 1{Z} e^{-\beta V\left(u_\mu T^{0\mu}-\mu J^-\right)}\ ,\quad u^\mu u_\mu=-1\ ,\ee
where $u^\mu$ constitutes the choice of a Lorentz frame, and $V$ is the volume of the system. The one-point functions
$\langle T^{\mu\nu}\rangle\ , \langle J^\mu\rangle$ are
functions of $\beta,\ u^\mu$ and $\mu$. For example, in the rest frame $u^\mu=(1,0,\cdots,0)$,
$ \vev{T^{\mu\nu}}= {\rm diag} (\ep_0, p_0, \cdots, p_0)$ and $\vev{J^\mu}=(n_0,0,\cdots,0)$
where $\varepsilon_0(\beta,\mu),\ p_0(\beta,\mu)$ and $n_0(\beta,\mu)$ are respectively the energy, pressure and charge densities. 
From now on for notional simplicity we shall drop the brackets from $\langle T^{\mu\nu}\rangle$ and $\langle J^\mu\rangle$. In a general frame $u^\mu$ they can be written as
\be T^{\mu\nu}=\varepsilon_0 u^\mu u^\nu+p_0\Delta^{\mu\nu}\ ,\quad J^\mu= n_0 u^\mu\ ,\ee
where $\Delta^{\mu\nu}$ is the projector transverse to the velocity,
\be \Delta^{\mu\nu}=\eta^{\mu\nu}+u^\mu u^\nu\ .\ee

Now consider a \emph{non-equilibrium} configuration where $T^{\mu\nu}$ and $J^\mu$ are slowly varying in space-time. More explicitly, if $L$ is the typical scale of variation of $T^{\mu\nu}$ and $J^\mu$, and $\ell$ is the typical microscopic relaxation scale
we have $L\gg \ell$. As discussed in Sec.~\ref{sec:1b}, each spacetime point can then be considered as in local equilibrium
defined by local values of the conserved quantities, or equivalently local values of $\b, \mu, u^\mu$. In other words, the system can be specified by $\beta(x),\ \mu(x)$ and $u^\mu(x)$. We can then write $T^{\mu\nu}$ and $J^\mu$ as
\begin{eqnarray}\notag
T^{\mu\nu}&=&\varepsilon_0 (x) u^\mu (x) u^\nu(x)+p_0(x)\Delta^{\mu\nu}(x)+\hat T^{\mu\nu}\ ,\\
J^\mu&=& n_0(x) u^\mu(x)+\hat J^\mu\label{eng1}\ ,\end{eqnarray}
where $\ep_0 (x) \equiv \ep_0 (\b (x), \mu (x))$ and similarly with $p_0 (x), n_0 (x)$. $\hat T^{\mu\nu}$ and $\hat J^\mu$
denote corrections from non-uniformity of $\beta(x), \mu (x), u^\mu (x)$. They can be expanded in terms of
the number of derivatives acting on $\beta,\mu,u^\mu$, with an effective expansion parameter $\ell \p_\mu \sim {\ell \ov L} \ll 1$.
 The hydrodynamical variables $\beta(x),\ \mu(x)$ and $u^\mu(x)$ constitute a set of $d+1$ unknowns, whose evolutions are determined by~(\ref{cons1}) and~(\ref{cons2}), which are $d+1$ equations. We thus have a closed set of dynamical equations.

The explicit expressions for $\hat T^{\mu\nu}$ and $\hat J^\mu$ are called \emph{constitutive relations}.
Naively, one just writes down the most general local expressions which are consistent with Lorentz symmetry. More explicitly, one finds to first derivative order (after using field redefinition freedom),
\begin{eqnarray}
\label{constrel1}\hat T^{\mu\nu}&=&-\eta\sigma^{\mu\nu}-\zeta \Delta^{\mu\nu}\theta+\cdots\\
\label{constrel2}\hat J^\mu&=& -\sigma T\Delta^{\mu\nu}\partial_\nu(\mu/T)+\chi_T\Delta^{\mu\nu}\partial_\nu T+\cdots
\end{eqnarray}
where $\p,\theta$ and $\sigma^{\mu\nu}$ are defined in (\ref{qcur4}), and  $\eta,\ \zeta,\ \sigma$ and $\chi_T$ are transport coefficients (they are functions of $\beta$ and $\mu$).
In~\eqref{constrel1}--\eqref{constrel2} one in fact gets more transport coefficients than desired; while $\eta, \ze, \sig$ corresponds respectively to
shear viscosity, bulk viscosity and conductivity, $\chi_T$ is not observed in nature.

So one needs to impose further constraints. A phenomenological constraint which appears to do the job is the local second law of thermodynamics: there exists an entropy current
\be \label{eng}
S^\mu=s_0(\beta(x),\mu(x))u^\mu(x)+\hat S^\mu\ ,
\ee
which upon using equations of motion~(\ref{cons1}) and (\ref{cons2}) satisfies
\be \label{sdiv}
\p_\mu S^\mu\geq 0\
\ee
order by order in derivative expansion. In~\eqref{eng}, $s_0(\beta,\mu)$ is the equilibrium entropy density, which is related to $\varepsilon_0, p_0$ and $n_0$ via standard thermodynamic relations, whereas $\hat S^\mu$ represents  derivative (i.e. non-equilibrium) corrections  to $S^\mu$. In practice, one writes down the most general local expression of $\hat S^\mu$
which are consistent with Lorentz symmetry, then sees whether it is possible to choose coefficients of $\hat S^\mu$
such that~\eqref{sdiv} is satisfied upon using~(\ref{cons1}) and (\ref{cons2}). One finds that this is only possible if
\be
\eta,\ \zeta,\ \sigma\geq 0\ ,\quad \chi_T=0\ .
\ee


One should also impose by hands the linear Onsager relations (from time reversal symmetry), i.e. the response matrix
for the external sources must be symmetric. The Onsager relations do not lead to any new constraints at the level of~\eqref{constrel1}--\eqref{constrel2}, but in general do at higher order in derivatives or more complicated systems (e.g. superfluids).

In summary, to obtain consistent hydrodynamic equations need to impose the following constraints:
\begin{enumerate}
\item The coefficients $\varepsilon_0,p_0, n_0, s_0$ are not independent, they satisfy the standard thermodynamic equilibrium relations. In other words, we need to impose local first law of thermodynamics.

\item Local second law of thermodynamics~\eqref{sdiv}.

\item Onsager relations.
\end{enumerate}
In the EFT approach discussed in Sec.~\ref{sec:eft2}--\ref{sec:hydroeft}, there is an action principle for obtaining the constitutive relations and all the above constraints are consequences of  the $Z_2$ dynamical KMS symmetry.

\section{A simple example of path integrals on CTP} \label{app:SK}

In this example we use a simple example to illustrate the role of boundary condition~\eqref{infbd} in generating couplings
between two segments of a CTP.

Consider the microscopic action for the harmonic oscillator
\be \label{actho} S=\frac 12\int_{t_i}^{t_f} dt(\dot x^2-\omega^2 x^2+xJ)\ ,\ee
where we included a linear coupling to the external source $J(t)$. Below we shall evaluate explicitly the generating functional (\ref{gen0}) with $\rho_0$ given by the vacuum state, i.e.
\be\rho_0=|\Omega\rangle\langle\Omega|\ .\ee
In order to do this, we break up the generating functional into a forward and a backward time evolutions,
\begin{eqnarray} \notag&&e^{W[J_1,J_2]}\\
\notag&&=\int d x_f\,\langle\Omega|U^\dag_{J_2}(t_f,t_i)| x_f\rangle\langle x_f|U_{J_1}(t_f,t_i)|\Omega\rangle\\
\notag&&=\int d x_f\,(\langle x_f|U_{J_2}(t_f,t_i)|\Omega\rangle)^*\langle x_f|U_{J_1}(t_f,t_i)|\Omega\rangle\ ,\\
&&\label{W12}\end{eqnarray}
where $U_J(t_f,t_i)$ is the evolution operator from $t_i$ to $t_f$ associated to the action (\ref{actho}), and $x(t_f)=x_f$.

Recall that for a harmonic oscillator the amplitude in going from position $x_i$ at $t=t_i$ to position $x_f$ at $t=t_f$ is 
\be\label{231} \langle x_f |U_J(t_f,t_i)| x_i \rangle= \mathcal N e^{\frac i\hbar \mathcal A[J,x_i,x_f]}\ ,\ee
where $\mathcal N$ is a constant which does not depend on $x_i,x_f$ (which we will suppress below), and 
\begin{eqnarray}\notag &&\mathcal A[J,x_i,x_f]\\
\notag&&=\frac 12 \int_{t_i}^{t_f} dt dt' J(t) G(t,t')J(t')+\frac 1{\sin\omega(t_f-t_i)}\,\times\\
\notag&&\times\int_{t_i}^{t_f}dt[x_i\sin\omega(t_f-t) +x_f\sin\omega(t-t_i)]J(t)\\
\notag&&+\frac{\omega}{2\sin\omega(t_f-t_i)} [(x_f^2+x_i^2)\cos\omega(t_f-t_i)-2x_f x_i]\ ,\\
&&\label{ma}
\end{eqnarray}
with
\begin{eqnarray}\notag G(t,t')=&&\frac{1} {2\omega\sin\omega(t_f-t_i)}\big(\cos\omega(t_f-t_i-|t-t'|)\\
&&-\cos\omega(t_i+t_f-t-t')\big)\ .\end{eqnarray}
Now in~\eqref{231}--(\ref{ma}) take the initial time $t_i=-\infty$, and use the standard trick to substitute $\omega\to \omega(1-i\ep)$, with an infinitesimal $\ep>0$. We then find that
\be
 \langle x_f|U_J(t_f,-\infty)| \Omega\rangle= e^{\frac i\hbar \mathcal A[J,x_f]},
 \ee
where
\begin{eqnarray}
\notag\mathcal A[J,x_f]&=& \frac 12 \int_{-\infty}^{t_f} dt dt' J(t) \bar G(t,t')J(t') \\
\notag&&+\int_{-\infty}^{t_f}dt\, x_f e^{\omega(i+\ep)(t-t_f)}J(t) +\frac i2\omega  x_f^2 \ ,\\
&&\label{ma1}\\
\notag\bar G(t,t')&=&\frac i{2\omega}(e^{-(i+\ep)\omega|t-t'|}-e^{(i+\ep)\omega(t+t'-2t_f)}) \\
&&\label{gb1} \ .
\end{eqnarray}

Eq.~(\ref{W12}) can then be written as
\be
\label{ma2}
e^{W[J_1,J_2]} 
=\int dx_f \, e^{\frac i\hbar (\mathcal A[J_1,x_f]- \mathcal A^*[J_2,x_f])} \ .
\ee
Integrating out $x_f$ in (\ref{ma2}) we find
\be
\begin{split} \label{012}
W[J_1,J_2]=&\frac 12 \int_{-\infty}^{t_f} dt dt' \, \big(J_1(t) \bar G(t,t')J_1(t') \\
&-J_2(t) \bar G^*(t,t')J_2(t')\big)\\
&+\frac i{4\omega}\bigg(\int dt(e^{\omega(i+\ep)(t-t_f)}J_1(t) \\
&-e^{\omega(-i+\ep)(t-t_f)}J_2(t))\bigg)^2 \ .
\end{split}\ee
First taking $\ep=0$
and then $t_f\to\infty$, Eq. (\ref{012}) becomes
\be\begin{split}\label{243} W[J_1,J_2]=&\frac 12 \int_{-\infty}^{\infty} dt dt' \bigg(J_1(t) G_F(t,t')J_1(t')\\
 &-J_2(t)  G_F^*(t,t')J_2(t')\\
 &-\frac i{2\omega}J_1(t)J_2(t')e^{i\omega(t-t')}\bigg),\end{split}\ee
where
\be G_F(t,t')=\frac i{2\omega}e^{-i\omega|t-t'|}.\ee
This illustrates that the boundary condition $x_1(t=t_f)=x_2(t=t_f)=x_f$ induces a coupling between $J_1$ and $J_2$ in the generating functional $W[J_1,J_2]$. Note that, had we taken $t_f\to\infty$ before taking $\ep\to 0$ in (\ref{012}), we would have obtained (\ref{243}) without the cross-term. This is because taking $t_f\to\infty$ with $\ep$ nonzero corresponds to putting the system in the ground state at $t_f=\infty$, leading to a decoupling of the two copies of the system.

\newpage
\section{Various discrete transformations} \label{app:c}

In this Appendix we list transformations of various tensors under
various discrete symmetries. They are important for obtaining the explicit forms of dynamical KMS transformations of various tensors in Sec.~\ref{sec:eft2}--\ref{sec:hydroeft}. For definiteness, we take $d=4$.
For notational simplicity we have suppressed the transformations of the arguments of all the functions, which are given in the first line of each table.

\medskip
\begin{widetext}
\begin{center}
\begin{tabular}{ |p{2cm}||p{3cm}|p{2cm}|p{2cm}|  }
 \hline
 \multicolumn{4}{|c|}{Discrete transformations in 3+1-dimension} \\
  \hline
 & $\sT$ &  $\sP \sT$ & $\sC \sP \sT$ \\
  \hline
 $x^\mu$   & $(-x^0,  x^i)$    & $- (x^0,  x^i)$    &  $- (x^0,  x^i)$ \\
 $u^\mu$   & $(u^0, - u^i)$    & $(u^0,  u^i)$    &  $(u^0,  u^i)$ \\
 $A_\mu$ &   $(A_0, - A_i)$     &  $(A_0,  A_i)$   &$- (A_0,  A_i)$\\
$\p_\mu$ &$(-\p_0,  \p_i)$ & $- (\p_0, \p_i)$&  $- (\p_0, \p_i)$ \\
 $\p = u^\mu \p_\mu$    & $- \p$  & $- \p$ &  $- \p$ \\
 $g_{\mu \nu}$ & $(g_{00}, - g_{0i}, g_{ij})$ & $ g_{\mu \nu}$
 & $g_{\mu  \nu}$  \\
  $\vp$ &    $- \vp$   &  $- \vp$  & $\vp$  \\
 \hline
\end{tabular}
\end{center}
\end{widetext}

\end{document}